\documentclass[english,reprint,onecolumn,prl,amsmath,amssymb, notitlepage]{revtex4-1}
\usepackage[T1]{fontenc}
\usepackage[latin9]{inputenc}
\usepackage{amsmath}
\usepackage{mathtools}
\usepackage{amsthm}
\usepackage{amssymb}
\usepackage{graphicx}
\usepackage{bm}
\usepackage[unicode=true, breaklinks=false, pdfborder={0 0 1}, backref=false, colorlinks=true, linkcolor=blue, urlcolor=blue, citecolor=blue,bookmarksopen=true]{hyperref}

\usepackage{braket}
\usepackage{bbm}

\usepackage[dvipsnames]{xcolor}

\allowdisplaybreaks
\newcommand{\tr}{\operatorname{tr}}
\newcommand{\id}{\mathbbm{1}}

\theoremstyle{plain}

\theoremstyle{plain}

\theoremstyle{plain}

\theoremstyle{plain}

\theoremstyle{definition}
\newtheorem{defn}{\protect\definitionname}

\let\oldsqrt\sqrt
\def\sqrt{\mathpalette\DHLhksqrt}
\def\DHLhksqrt#1#2{%
\setbox0=\hbox{$#1\oldsqrt{#2\,}$}\dimen0=\ht0
\advance\dimen0-0.2\ht0
\setbox2=\hbox{\vrule height\ht0 depth -\dimen0}%
{\box0\lower0.4pt\box2}}

\let\originalleft\left
\let\originalright\right
\renewcommand{\left}{\mathopen{}\mathclose\bgroup\originalleft}
\renewcommand{\right}{\aftergroup\egroup\originalright}

\renewcommand\bra[1]{{\langle{#1}|}}
\makeatletter
\renewcommand\ket[1]{%
  \@ifnextchar\bra{\k@t{#1}\!}{\k@t{#1}}%
}
\newcommand\k@t[1]{{|{#1}\rangle}}
\makeatother

\usepackage{babel}
\providecommand{\corollaryname}{Corollary}
\providecommand{\lemmaname}{Lemma}
\providecommand{\propositionname}{Proposition}
\providecommand{\theoremname}{Theorem}
\providecommand{\definitionname}{Definition}

\makeatother

\begin{document}

\title[Witnessing non-objectivity]{Witnessing non-objectivity in the framework of Strong Quantum Darwinism}

\author{Thao P. Le}
\email{thao.le.16@ucl.ac.uk}

\affiliation{Dept. of Physics and Astronomy, University College London, Gower Street, London WC1E 6BT}

\author{Alexandra Olaya-Castro}
\email{a.olaya@ucl.ac.uk}
\affiliation{Dept. of Physics and Astronomy, University College London, Gower Street, London WC1E 6BT}

\begin{abstract}
Quantum Darwinism is a compelling theory that describes the quantum-to classical transition as the emergence of objectivity of quantum systems. Spectrum broadcast structure and strong quantum Darwinism are two extensions of this theory with emphasis on state structure and information respectively. The complete experimental verification of these three frameworks, however, requires quantum state tomography over both the system and accessible environments, thus limiting the feasibility and scalability of experimental tests. Here, we introduce a subspace-dependent objectivity operation and construct a witness that detects non-objectivity by comparing the dynamics of the system-environment state with and without the objectivity operation. We then propose a photonic experimental simulation that implements the witnessing scheme. Our work proposes a route to further experimental exploration of the quantum to classical transition.
\end{abstract}

\maketitle

\section{Introduction}

Everyday macroscopic systems are \emph{objective} in the sense that certain information about their states are, in principle, knowable by multiple independent observers. In contrast, quantum systems are subjective in the sense that they can be measured by in multiple different bases, causing different observers to obtain different information about the system. From this perspective, the quantum-to-classical transition is then understood as the emergence of objective states as a system interacts with its surrounding environment \citep{Zurek2009, Horodecki2015,Le2019}.

Quantum Darwinism proposed by Zurek \citep{Zurek2009} describes how objectivity can emerge from microscopic quantum behaviour: as systems decohere via interactions with their environment, specific information about their state can be duplicated into multiple parts of the environment. \emph{Information} is ``Darwinistic'' as certain classical information tends to proliferate to the detriment of other types of information. Strong quantum Darwinism \citep{Le2019}, which we have proposed, goes a step forward by identifying the necessary and sufficient conditions for which objectivity emerges and by establishing the formal mathematical equivalence between analysing objectivity through an information approach and through the state broadcasting  structure \citep{Horodecki2015}. Strong quantum Darwinism therefore provides a solid framework to bridge different levels of objectivity and the appearance of classically. Our overall goal with this paper is thus to motivate further theoretical and experimental scrutiny of our proposed framework.

The importance of the concept of (non) objectivity for a physical system can be appreciated by considering incoherent systems. We typically understand incoherent systems to be classical as  explained by decoherence theory \citep{Schlosshauer2014}. However, decoherence theory does not resolve the quantum-to-classical transition as the lack of quantum coherence is not sufficient for objectivity. For example, a multipartite maximally mixed state is considered non-objective, as there is no information\textemdash neither quantum nor classical\textemdash shared between the systems. In general, incoherent states do not have perfect correlations and thus are also non-objective. This leads to the concept of \emph{classical non-objectivity}, which could be particularly relevant for complex systems operating at the quantum-classical boundary \cite{Arndt2009, OReilly2014}.
Those systems could be incoherent yet non-objective, and their non-objectivity could lead to particular advantages for a system operating in such a regime.

While the theoretical exploration of quantum Darwinism has been significant \citep{Zurek2009,Brandao2015,Knott2018,Korbicz2017,Qi2020, Balaneskovic2015,Balaneskovic2016,Ollivier2004,Blume-Kohout2005,Riedel2012,Zwolak2009,Zwolak2010,Zwolak2016,Giorgi2015,Zwolak2014,Zwolak2017,Lampo2017,Riedel2010,Riedel2011,Korbicz2014,Horodecki2015,Perez2010,Le2018,Korbicz2017a,Tuziemski2018,Blume-Kohout2006,Roszak2019,Campbell2019,Le2018,Garcia-Perez2019,Tuziemski2019,Oliveira2019, Roszak2020, Milazzo2019, Lorenzo2020a,Le2020,Tuziemski2020}, experimental investigations of this theory are scarce, in part hindered by the fact that complete studies need quantum state tomography of the system and the accessible environments.
Previously, a number of experiments on open quantum dots suggested indirect links between particular transport properties and quantum Darwinism \citep{Brunner2008,Burke2010,Ferry2011,Brunner2012,Brunner2010,Ferry2015}, but there was no quantitative characterisation of quantum Darwinism.  More recently, advances in the simulation of open quantum systems with photonic setups and the sophisticated control of spin systems has changed the landscape for experimental investigations of the quantum Darwinism. Specifically, three experiments explore the emergency of objectivity in photonic cluster states \citep{Ciampini2018}, photonic quantum simulators \citep{Chen2019} and nitrogen vacancy centers \citep{Unden2019}. In the two photonic experiments \citep{Ciampini2018,Chen2019}, full quantum state tomography of the simulated system and environments is used to determine the state and hence characterise their (non-)objectivity.The nitrogen vacancy experiment \citep{Unden2019} is remarkable as it explores quantum Darwinism in a matter system and therefore brings us closer to understand the quantum to classical transition for realistic open quantum systems. However, this work has the caveat that it determines only the classical Holevo information; therefore, is technically not sufficient for asserting any of Zurek's quantum Darwinism \citep{Zurek2009}, strong quantum Darwinism \citep{Le2019}, nor spectrum broadcast structure \citep{Horodecki2015} which require extra information such as the quantum discord. 

Quantum state tomography is required to fully characterise quantum Darwinism. This hinders the scale and scope of experiments possible, especially of those with larger and more realistic environments, which has led to work on more efficient tomography schemes given some allowable error \cite{Cramer2010,ODonnell2015,Haah2017,Aaronson2018}. This problem is closely linked to the difficulties in characterising quantum entanglement and other quantum correlations \citep{Guehne2009,Adesso2016}. One solution has been entanglement witnesses: operators and schemes which detect nonclassical correlations much more simply: however, a single witness alone is not capable of detecting all entangled states \citep{Lewenstein2000}. The most famous witness are the Bell inequalities \citep{Bell1964}\textemdash mathematical inequalities that are satisfied by a classical theory but can be broken under some forms of quantum entanglement \citep{Terhal2000}.

As a possible pathway for experimental testing of Quantum Darwinism, both in its strong form and in larger system-environments where state tomography becomes intractable,  we introduce a non-objectivity witness in the strong quantum Darwinism framework. We consider a \emph{preferred} subspace in which objectivity could occur, analogous to the preferred basis of quantum coherence. Our scheme detects non-objectivity by comparing the evolution of the system-environment state with and without objectivity-enforcing operations. To motivate experimental tests of our scheme, we present an experimental quantum photonic simulation that follows our witnessing scheme. By comparing the number of measurements needed for state tomography versus those needed for the proposed witness, we show that, with sufficiently good components, our scheme provides a significant advantage. Thus, the witness scheme we present can further advance the experimental testing of Quantum Darwinism, and the understanding the quantum-to-classical transition in terms of (non) objectivity.

The paper is organised as follows: in Sec. \ref{sec:Frameworks-of-Quantum-Darwinism} we define objectivity and review the various mathematical frameworks of quantum Darwinism. In Sec. \ref{sec:Witnessing-non-objectivity} we describe the non-objective witness scheme. In Sec. \ref{sec:Photonic-Experiment-Proposal} we propose a quantum photonic experiment and provide numerical simulation results. We end with a discussion in Sec. \ref{sec:Discussion}.

\section{Frameworks of Quantum Darwinism\label{sec:Frameworks-of-Quantum-Darwinism}}

Suppose we have a system $\mathcal{S}$, numerous environments $\left\{ \mathcal{E}_{k}\right\} _{k}$, and hypothetical observers each with access to separate environments. Quantum Darwinism describes how the spread of information leads to system objectivity: $\mathcal{S}$ becomes objective when the environments $\mathcal{E}_{k}$ contain full information about the system state.

\begin{defn}
Objectivity \cite{Ollivier2004,Zurek2009,Horodecki2015}:
A system state is \emph{objective} if it is (1) simultaneously accessible to many observers (2) who can all determine the state independently \emph{without perturbing it} and (3) all arrive at the same result .

\end{defn}

There are three mathematical frameworks that characterise the objectivity condition precisely: Zurek's quantum Darwinism \citep{Zurek2009}, strong quantum Darwinism \citep{Le2019} and spectrum broadcast structure \citep{Horodecki2015}. In this section, we will give these conditions, before focusing on strong quantum Darwinism.

\begin{defn}
Zurek's quantum Darwinism (\textbf{QD}) \citep{Zurek2009}: Here, objectivity is said to occur when the quantum mutual information between system and environment is equal to the information contained in the system:
\begin{equation}
I\left(\mathcal{S}:\mathcal{E}_{k}\right)=H\left(\mathcal{S}\right),\label{eq:QD_condition}
\end{equation}
where $I\left(\mathcal{S}:\mathcal{E}_{k}\right)=H\left(\mathcal{S}\right)+H\left(\mathcal{\mathcal{E}}_{k}\right)-H\left(\mathcal{S\mathcal{E}}_{k}\right)$ is the quantum mutual information between system and environment $\mathcal{E}_{k}$, and $H\left(\mathcal{S}\right)=-\tr\rho_{\mathcal{S}}\log\rho_{\mathcal{S}}$ is the von Neumann entropy of the system with reduced state $\rho_{\mathcal{S}}$.
\end{defn}

\begin{defn}
Strong Quantum Darwinism (\textbf{SQD}) \citep{Le2019}: Here, objectivity occurs when the quantum mutual information is equal to the classical information shared between the system and environment
\begin{equation}
I\left(\mathcal{S}:\mathcal{E}_{k}\right)=\chi\left(\mathcal{S}:\mathcal{E}_{k}\right) =H\left(\mathcal{S}\right),\qquad\mathcal{D}\left(\mathcal{S}:\mathcal{E}_{k}\right)=0,\label{eq:SQD}
\end{equation}
where $\chi\left(\mathcal{S}:\mathcal{E}_{k}\right)$ is the classical accessible information given by the Holevo quantity. The quantum discord is as follows:
\begin{equation}
\mathcal{D}\left(\mathcal{S}:\mathcal{E}_{k}\right)=\min_{\left\{ \Pi_{\mathcal{S}}\right\} }\sum_{i}p_{i}H\left(\rho_{\mathcal{E}_{k}|i}\right)+H\left(\rho_{S}\right)-H\left(\rho_{\mathcal{SE}_{k}}\right),
\end{equation}
where $\rho_{\mathcal{E}_{k}|i}$ is the conditional state on subenvironment $\mathcal{E}_{k}$ after measurement result $i$ on $\mathcal{S}$, using the positive-operator-valued measure (POVM) $\left\{ \Pi_{\mathcal{S}}\right\} $ \citep{Henderson2001,Ollivier2001}. In SQD, the quantum discord\textemdash genuinely quantum correlations\textemdash must be zero. The Holevo information can then be defined as the difference with the quantum mutual information \citep{Zwolak2013}:
\begin{equation}
\chi\left(\mathcal{S}:\mathcal{E}_{k}\right)  =I\left(\mathcal{S}:\mathcal{E}_{k}\right)-\mathcal{D}\left(\mathcal{S}:\mathcal{E}_{k}\right).
\end{equation}
\end{defn}

\begin{defn}
Spectrum broadcast structure (\textbf{SBS}) \citep{Horodecki2015}: Here, objectivity occurs when the system-environment state has the following structure:
\begin{equation}
\rho_{\mathcal{S}\mathcal{E}_{1}\cdots\mathcal{E}_{N}}=\sum_{i}p_{i}\ket{i}\bra{i}_{\mathcal{S}}\otimes\rho_{\mathcal{E}_{1}|i}\otimes\cdots\otimes\rho_{\mathcal{E}_{N}|i},\qquad\rho_{\mathcal{E}_{k}|i}\rho_{\mathcal{E}_{k}|j}=0\quad\forall i\neq j,\quad\forall k,\label{eq:SBS}
\end{equation}
where the conditional environment states $\rho_{\mathcal{E}_{k}|i}$ are perfectly distinguishable.
\end{defn}
The perfect distinguishability allows any observer with access to $\mathcal{E}_{k}$ to construct a measurement that will perfectly determine the index $i$.

Spectrum broadcast structure differs from strong quantum Darwinism with the condition of strong independence, where, conditioned on the system, the subenvironments share no (extra) correlations amongst each other \citep{Le2019}.

\begin{figure}
    \centering
    \includegraphics[width=1\textwidth]{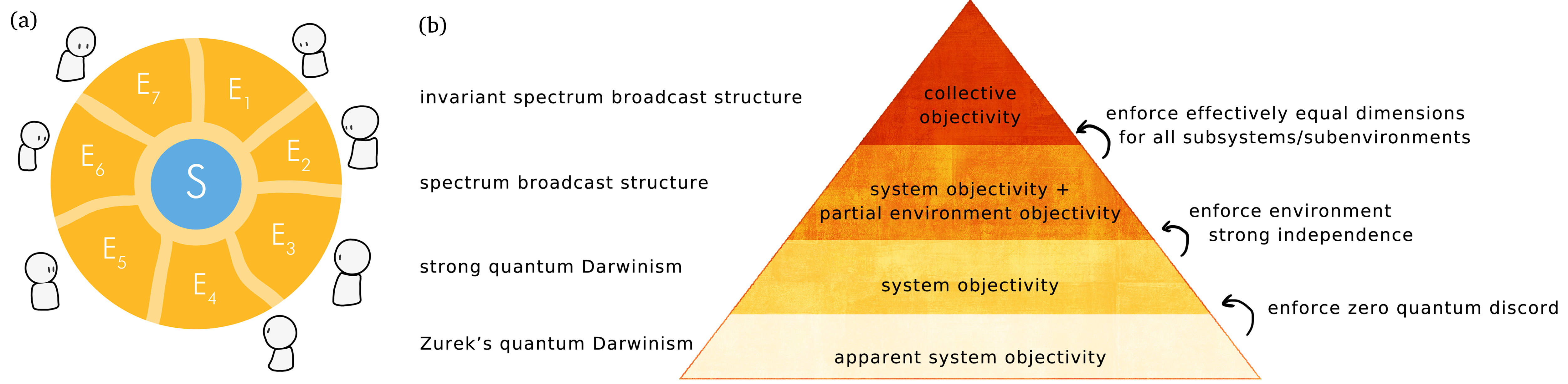}
    \caption{(a) Depiction of Quantum Darwinism: the central system interacts with its surrounding environments. Observers in turn measure subenvironments in order to determine information about the system. (b) Overview of the different frames of Quantum Darwinism: Zurek's quantum Darwinism \citep{Zurek2009}, strong quantum Darwinism \citep{Le2019}, spectrum broadcast structure \citep{Horodecki2015}, and invariant spectrum broadcast structure (Appendix \ref{app:Invariant-Spectrum-Broadcast}).}
    \label{fig:QD_overview}
\end{figure}

These three frameworks correspond to different levels of objectivity, which we summarise in Fig.~\ref{fig:QD_overview}: Zurek's quantum Darwinism describes apparent objectivity, whilst spectrum broadcast structure describes system objectivity with partial environment objectivity. In contrast, Strong quantum Darwinism describes the precise minimal conditions for objectivity of the system. Thus, in this paper, we will focus on strong quantum Darwinism.

\subsection{Subspace-dependent Strong Quantum Darwinism\label{subsec:Subspace-dependent-Strong-Quantum-Darwinism}}

Here, we introduce the notion of subspace-dependent strong quantum Darwinism, upon which we will build our witness in Sec. \ref{sec:Witnessing-non-objectivity}. Environments and systems tend to have a limited set of bases in which we routinely measure. Furthermore, states that subspace-dependent SQD form a convex set, and hence are closed under convex combinations, similar to the convexity of separable states and incoherent states. Subspace-dependent SQD is built from basis-dependent discord \cite{Egloff2018,Yadin2016}, which in turn has strong connections with quantum coherence \citep{Yadin2016, Streltsov2017}.

Strong quantum Darwinism is equivalent to \emph{bipartite} spectrum broadcast structure \citep{Le2019}. That is, if the state on $\mathcal{SF}=\mathcal{SE}_{1}\cdots\mathcal{E}_{F}$ has strong quantum Darwinism, then the system $\mathcal{S}$ with the full fragment $\mathcal{F}$ has bipartite spectrum broadcast structure:
\begin{equation}
\rho_{\mathcal{SF}}  =\sum_{i}p_{i}\ket{i}\bra{i}\otimes\rho_{\mathcal{F}|i},\qquad\rho_{\mathcal{F}|i}\rho_{\mathcal{F}|j}=0\quad\forall i\neq j,\label{eq:bipartiteSBSF}
\end{equation}
and the system $\mathcal{S}$ with the individual components also has  bipartite spectrum broadcast structure:
\begin{equation}
\rho_{\mathcal{SE}_{k}}  =\sum_{i}p_{i}\ket{i}\bra{i}\otimes\rho_{\mathcal{E}_{k}|i},\qquad\rho_{\mathcal{E}_{k}|i}\rho_{\mathcal{E}_{k}|j}=0\quad\forall i\neq j,\text{ for each }k=1,\ldots,F.\label{eq:bipartiteSBSk}
\end{equation}
As the different conditional fragment states $\left\{ \rho_{\mathcal{F}|i}\right\}_{i}$ are orthogonal, we can define disjoint subspaces $\left\{ \mathcal{H}_{\mathcal{F}|i}\right\}_{i}$ in which they lie. Similarly, the different conditional sub-environment states $\left\{ \rho_{\mathcal{E}_{k}|i}\right\} _{i}$ are orthogonal and we can also define the disjoint subspaces $\left\{ \mathcal{H}_{\mathcal{E}_{k}|i}\right\} _{i}$. Due to the state structure above, the conditional disjoint subspace of $\mathcal{F}$ is the tensor product of the conditional disjoint subspaces in $\mathcal{E}_{k}$:
\begin{equation}
\mathcal{H}_{\mathcal{F}|i}=\mathcal{H}_{\mathcal{E}_{1}|i}\otimes\mathcal{H}_{\mathcal{E}_{2}|i}\otimes\cdots\otimes\mathcal{H}_{\mathcal{E}_{F}|i}.\label{eq:tensor_of_hilbert_spaces}
\end{equation}
Let $\Pi_{\mathcal{E}_{k}|i}$ be the projector into the subspace $\mathcal{H}_{\mathcal{E}_{k}|i}$. The projector onto the tensor product space $\mathcal{H}_{\mathcal{F}|i}$ is simply
\begin{equation}
\Pi_{\mathcal{F}|i}  =\Pi_{\mathcal{E}_{1}|i}\otimes\cdots\otimes\Pi_{\mathcal{E}_{F}|i}.\label{eq:Fragment_projection}
\end{equation}
The action of this projector $\Pi_{\mathcal{F}|i}\rho_{\mathcal{F}}\Pi_{\mathcal{F}|i}$ preserves some correlations between the environment states in general as the projectors $\Pi_{\mathcal{E}_{k}|i}$ can have rank greater than one. This is allowed within the framework of strong quantum Darwinism. In contrast, these correlations are not allowed in spectrum broadcast structure due to strong independence \cite{Horodecki2015}.

In subspace-dependent strong quantum Darwinism, we define the preferred basis $\left\{ \ket{i}_{\mathcal{S}}\right\} _{i}$ on the system, and we define the preferred objective subspace partitioning for the environments, which we encode in the projectors $\left\{ \Pi_{\mathcal{F}|i}\right\} $ from Eq.~(\ref{eq:Fragment_projection}). The following \emph{objectivity operation} projects any input state into an incoherent state satisfying strong quantum Darwinism, in the fixed subspaces as defined:
\begin{equation}
\Gamma_{\mathcal{SF}}^{\text{SQD}}\left(\rho\right)=\sum_{i}\left(\ket{i}\bra{i}_{\mathcal{S}}\otimes\Pi_{\mathcal{F}|i}\otimes\id_{\mathcal{E}\backslash\mathcal{F}}\right)\rho\left(\ket{i}\bra{i}_{\mathcal{S}}\otimes\Pi_{\mathcal{F}|i}\otimes\id_{\mathcal{E}\backslash\mathcal{F}}\right).\label{eq:SQD_objective_op}
\end{equation}
This is comparable to the discord-breaking measurement in Ref. \citep{Gessner2013}, or the decoherence operation in Ref. \citep{Knee2018}.

From this, we can define a measure of subspace-dependent non-objectivity using the trace norm distance:
\begin{equation}
M^{\text{SQD}}\left(\rho_{\mathcal{SF}}\left(t\right)\right)=\left\Vert \rho_{\mathcal{SF}}\left(t\right)-\Gamma_{\mathcal{SF}}^{\text{SQD}}\left(\rho_{\mathcal{SF}}\left(t\right)\right)\right\Vert _{1}.\label{eq:measureSQD}
\end{equation}
The maximum value of the measure is $\max_{\rho_{\mathcal{SF}}\left(t\right)}M^{\text{SQD}}\left(\rho_{\mathcal{SF}}\left(t\right)\right)=1$ and occurs when the system-fragment is completely nonobjective. The typical trace-norm distance $\left\Vert \rho-\sigma\right\Vert _{1}$
upper-bound (of two) occurs when $\rho$ and $\sigma$ have orthogonal support. However, due to the nature of the objectivity operation,  orthogonality between the two terms only occurs when $\Gamma_{\mathcal{SF}}^{\text{SQD}}\left(\rho_{\mathcal{SF}}\left(t\right)\right)=0$.

In the following section, we will employ the objectivity operation to create a witness that lower bounds the value of the measure in Eq.~(\ref{eq:measureSQD}).

\section{Witnessing non-objectivity\label{sec:Witnessing-non-objectivity}}

\begin{figure}
\begin{centering}
\includegraphics[width=0.6\textwidth]{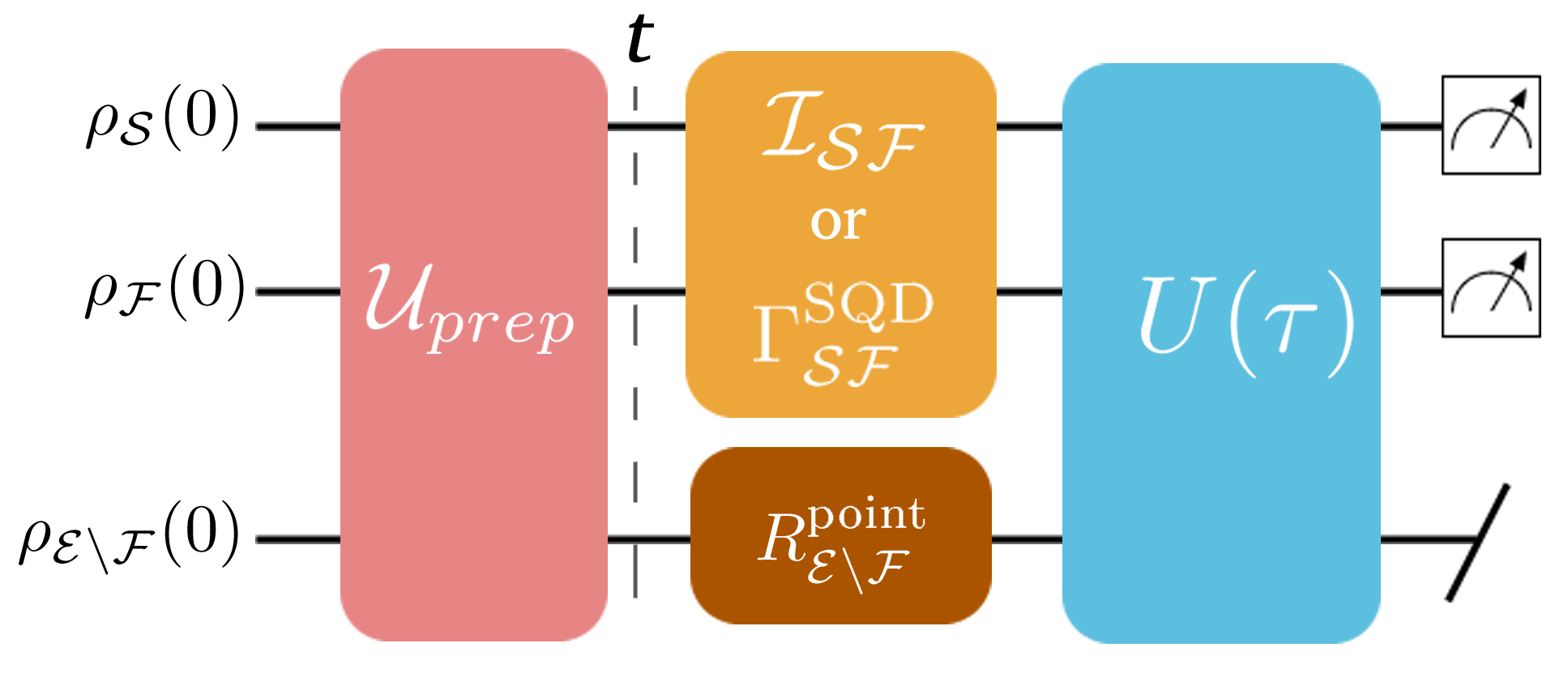}
\par\end{centering}
\caption{\textbf{Protocol for the non-objectivity witness}. The system-environment is prepared into some state  $\rho_{\mathcal{SE}}\left(t\right)=\mathcal{U}_{prep}\left(\rho_{\mathcal{SE}}\left(0\right)\right)$. A point channel $R_{\mathcal{E}\backslash\mathcal{F}}^{\text{point}}$ [Eq.~(\ref{eq:point_channel})] is applied on the ``un-accessed'' environment $\mathcal{E}\backslash\mathcal{F}$ to ensure that the witness does not detect extraneous correlations. We can either leave the system-fragment $\mathcal{SF}$ untouched (identity channel $\mathcal{I_{\mathcal{SF}}}$) or apply the objectivity operation $\Gamma_{\mathcal{SF}}^{\text{SQD}}$ [Eq.~(\ref{eq:SQD_objective_op})]. The system-environment then undergoes some unitary evolution $U\left(\tau\right)$, and is measured at the end of the protocol. The witness comes from the comparison between the final state with or without the objectivity operation.\label{fig:Protocol-for-a-witness}}
\end{figure}

We will construct a witness to detect non-objectivity between a system and some collection of subenvironments. A non-zero witness implies non-objectivity relative to the pre-defined basis and subspaces. Our scheme is illustrated in Fig.~\ref{fig:Protocol-for-a-witness}: by measuring the difference between two alternative system-environment evolutions, we can determine the amount of non-objectivity present. This method in the spirit of previous schemes for witnessing quantum discord \citep{Gessner2013} and quantum coherence \citep{Knee2018}. We could alternatively use a combination of a discord witness \citep{Gessner2013, Yu2011, Rahimi2010, Maziero2012, Zhang2011} combined with a measurement of the classical information. However, unlike the scheme we will introduce below, this will not produce a single witness value for non-objectivity, as non-objectivity scales as $\mathcal{D} - \chi$  \cite{Le2019} i.e.  with opposing dependence on quantum and classical information whilst experimental witnesses tend to lower-bound the information. Another possibility is to use a traditional witness operator $W$ leading to values $\tr[W\rho]$.  However, we expect this to require multiple joint copies of the state at the same time, e.g. $\rho_{\mathcal{SE}}^{\otimes 4}$ in the discord witness of Ref. \citep{Yu2011}, which would make the experimental scheme more cumbersome given the already large dimensions of the environment.

The system-environment evolution proceeds as follows: At time $t=0$, the system and environment starts out in joint initial state state, $\rho_{\mathcal{SE}}\left(0\right)$. The full system-environment then evolves under the action of a unitary $U\left(t\right)$ such that the state at time $t$ is
\begin{equation}
\rho_{\mathcal{SE}}\left(t\right)=U\left(t\right)\rho_{\mathcal{SE}}\left(0\right)U^{\dagger}\left(t\right)=\mathcal{U}_{t}\left[\rho_{\mathcal{SE}}\left(0\right)\right].
\end{equation}
Our goal is to witness non-objectivity of the system and fragment $\rho_{\mathcal{SF}}\left(t\right)=\tr_{\mathcal{E}\backslash\mathcal{F}}\left[\rho_{\mathcal{SE}}\left(t\right)\right]$.

To do so, we first must ensure that the witness does not pick up on extraneous correlations between the observed system-fragment $\mathcal{SF}$ and the rest of the environment $\mathcal{E}\backslash \mathcal{F}$. We use a point channel on the remainder environment, which discards the $\mathcal{E}_{F+1}\ldots\mathcal{E}_{N}$ states and prepares a new (uncorrelated) state. 
\begin{equation}
R_{\mathcal{E}\backslash\mathcal{F}}^{\text{point}}\left(\rho_{\mathcal{S}\mathcal{E}_{1}\cdots\mathcal{E}_{N}}\left(t\right)\right)  =\tr_{\mathcal{E}\backslash\mathcal{F}}\left[\rho_{\mathcal{S}\mathcal{E}_{1}\cdots\mathcal{E}_{N}}\left(t\right)\right]\otimes\rho_{\mathcal{E}\backslash\mathcal{F}}^{\text{new}}=\rho_{\mathcal{SF}}\left(t\right)\otimes\rho_{\mathcal{E}\backslash\mathcal{F}}^{\text{new}}.\label{eq:point_channel}
\end{equation}
Note that the point channel is crucial to isolate the correlations we want to test. For example, the authors of Ref. \citep{Knee2018} profess an ambiguity in one of their witnesses\textemdash as to whether it is detecting system coherences or system-environment correlations. This ambiguity would be removed with the addition of a correlation breaking channel, as we have done here.

If the point channel is the only operation that we enact at time $t$, then the system-environment subsequently evolves under some unitary evolution $U\left(\tau\right)$:
\begin{equation}
\rho_{\mathcal{SE}}\left(t+\tau\right)=\mathcal{U}_{\tau}\circ\left(\id_{\mathcal{SF}}\otimes R_{\mathcal{E}\backslash\mathcal{F}}^{\text{point}}\right)\left[\rho_{\mathcal{SE}}\left(t\right)\right]=\mathcal{U}_{\tau}\left[\rho_{\mathcal{SF}}\left(t\right)\otimes\rho_{\mathcal{E}\backslash\mathcal{F}}^{\text{new}}\right].
\end{equation}
Finally, we conduct a measurement $M_{\mathcal{S}\mathcal{E}}$, giving us the probability:
\begin{equation}
P_{\id_{\mathcal{SF}}}=\tr\left[M_{\mathcal{SE}}\rho_{\mathcal{SE}}\left(t+\tau\right)\right].\label{eq:P_id}
\end{equation}
One possible measurement operator $M_{\mathcal{SE}}$ is simply the projector onto zero: $M_{\mathcal{SE}}^{\text{e.g.}}=\ket{0}\bra{0}\otimes\ket{0}\bra{0}\otimes\cdots\otimes\ket{0}\bra{0}$.

In an alternative evolution, we could \emph{also} apply the objectivity operation from Eq.~(\ref{eq:SQD_objective_op}) on the system-fragment at time $t$ in conjunction with the point channel such that the subsequent final state is:
\begin{equation}
\rho_{\mathcal{SE}}^{\prime}\left(t+\tau\right)  =\mathcal{U}_{\tau}\left[\Gamma_{\mathcal{SF}}^{\text{SQD}}\left(\rho_{\mathcal{SF}}\left(t\right)\right)\otimes\rho_{\mathcal{E}\backslash\mathcal{F}}^{\text{new}}\right],
\end{equation}
leading to the alternative probability of measurement $M_{\mathcal{SE}}$ at time $t+\tau$:
\begin{equation}
P_{\Gamma_{\mathcal{SF}}^{\text{SQD}}}=\tr\left[M_{\mathcal{SE}}\rho_{\mathcal{SE}}^{\prime}\left(t+\tau\right)\right].\label{eq:P_obj}
\end{equation}

The absolute difference between these probabilities is our witness for non-objectivity:
\begin{equation}
W^{\text{SQD}}\left(M_{\mathcal{SE}}\right)  =\left|P_{\id_{\mathcal{SF}}}-P_{\Gamma_{\mathcal{SF}}^{\text{SQD}}}\right| =\left|\tr\left[M_{\mathcal{SE}}\mathcal{U}_{\tau}\left[\left\{ \rho_{\mathcal{SF}}\left(t\right)-\Gamma_{\mathcal{SF}}^{\text{SQD}}\left(\rho_{\mathcal{SF}}\left(t\right)\right)\right\} \otimes\rho_{\mathcal{E}\backslash\mathcal{F}}^{\text{new}}\right]\right]\right|.
\end{equation}

The witness lower bounds the non-objectivity measure from Eq.~(\ref{eq:measureSQD}):
\begin{equation}
W^{SQD}\left(M_{\mathcal{SE}}\right)\leq M^{SQD}\left(\rho_{\mathcal{SF}}\left(t\right)\right).\label{eq:witness_lowerbounds_measure}
\end{equation}
This can be shown as follows: if we maximised over measurement operators in the witness, then
\begin{equation}
W^{SQD}\left(M_{\mathcal{SE}}\right)\leq\max_{M_{\mathcal{SE}}}W^{SQD}\left(M_{\mathcal{SE}}\right)  =\max_{M_{\mathcal{SE}}^{\prime}}\left|\tr\left[M_{\mathcal{SE}}^{\prime}\left(\left\{ \rho_{\mathcal{SF}}\left(t\right)-\Gamma_{\mathcal{SF}}^{SQD}\left(\rho_{\mathcal{SF}}\left(t\right)\right)\right\} \otimes\rho_{\mathcal{E}\backslash\mathcal{F}}^{\text{new}}\right)\right]\right|,
\end{equation}
where $M_{\mathcal{SE}}^{\prime}=U^{\dagger}\left(\tau\right)M_{\mathcal{SE}}U\left(\tau\right)$ satisfies $\left\Vert M_{\mathcal{SE}}^{\prime}\right\Vert =\max\left\{ \left|\lambda_{i}\right|:\lambda_{i}\text{ eigenvalue of }M_{\mathcal{SE}}^{\prime}\right\} \leq1$. Thus
\begin{eqnarray}
\max_{M_{\mathcal{SE}}}W^{SQD}\left(M_{\mathcal{SE}}\right) & \leq&\sup_{\left\Vert B\right\Vert \leq1}\left|\tr\left[B\left(\left\{ \rho_{\mathcal{SF}}\left(t\right)-\Gamma_{\mathcal{SF}}^{SQD}\left(\rho_{\mathcal{SF}}\left(t\right)\right)\right\} \otimes\rho_{\mathcal{E}\backslash\mathcal{F}}^{\text{new}}\right)\right]\right|\label{eq:max_to_sup}\\
 & =&\left\Vert \left\{ \rho_{\mathcal{SF}}\left(t\right)-\Gamma_{\mathcal{SF}}^{SQD}\left(\rho_{\mathcal{SF}}\left(t\right)\right)\right\} \otimes\rho_{\mathcal{E}\backslash\mathcal{F}}^{\text{new}}\right\Vert _{1}\\
 & =&\left\Vert \rho_{\mathcal{SF}}\left(t\right)-\Gamma_{\mathcal{SF}}^{SQD}\left(\rho_{\mathcal{SF}}\left(t\right)\right)\right\Vert _{1}  = M^{SQD}\left(\rho_{\mathcal{SF}}\left(t\right)\right),
\end{eqnarray}
where the trace norm of the measure can be written as supremum over Hermitian operators $B$ where $\left\Vert B\right\Vert \leq1$ ($-I\leq B\leq I$), and where $\left\Vert A\otimes B\right\Vert _{1}=\left\Vert A\right\Vert _{1}\cdot\left\Vert B\right\Vert _{1}$. This gives Eq.~(\ref{eq:witness_lowerbounds_measure}).

The witness depends on suitable choices of the unitary operation $\mathcal{U}_{\tau}$ and final measurement $M_{\mathcal{SE}}$ to ensure that the two alternative probabilities [Eqs.~(\ref{eq:P_id}) and (\ref{eq:P_obj})] are different. The second unitary in particular, $U(\tau)$ can be chosen to pick up correlations that will otherwise go undetected \cite{Knee2018} by effectively changing the basis of the final measurements, especially for nonlocal unitaries that can then implement nonlocal measurements.

Like coherence and its witnesses, the witness we have presented is system-basis and environment-subspace dependent. Quantum Darwinism, in its fullest, requires optimisation over all bases. However, we wish for a scheme less intensive than full quantum state tomography. Furthermore, realistically, we are constrained in what we can measure, especially when it comes to environments: often, there are only a limited number of degrees of freedom with encoded information that we can access, and/or only a limited number of degrees of freedom that can \emph{possibly} encode information about the system of interest. As such, there should be a naturally preferred basis and subspace. 

\section{Quantum Photonic Simulation Experimental Proposal \label{sec:Photonic-Experiment-Proposal}}

In this section, we propose an experiment to apply the witness onto a system and environment composed of photons with information encoded in the polarisation degrees of freedom. Our scheme is particularly suited to photonic qubits over spin qubits and other related systems using magnetic resonance as the objectivity operation from Eq.~(\ref{eq:SQD_objective_op}) relies on projective measurement. We will first present the setup, then numerical simulations, and followed by a comparison between the non-objectivity witness and quantum state tomography.

\subsection{Overall Setup}

\begin{figure}
\begin{centering}
\includegraphics[width=0.36\textwidth]{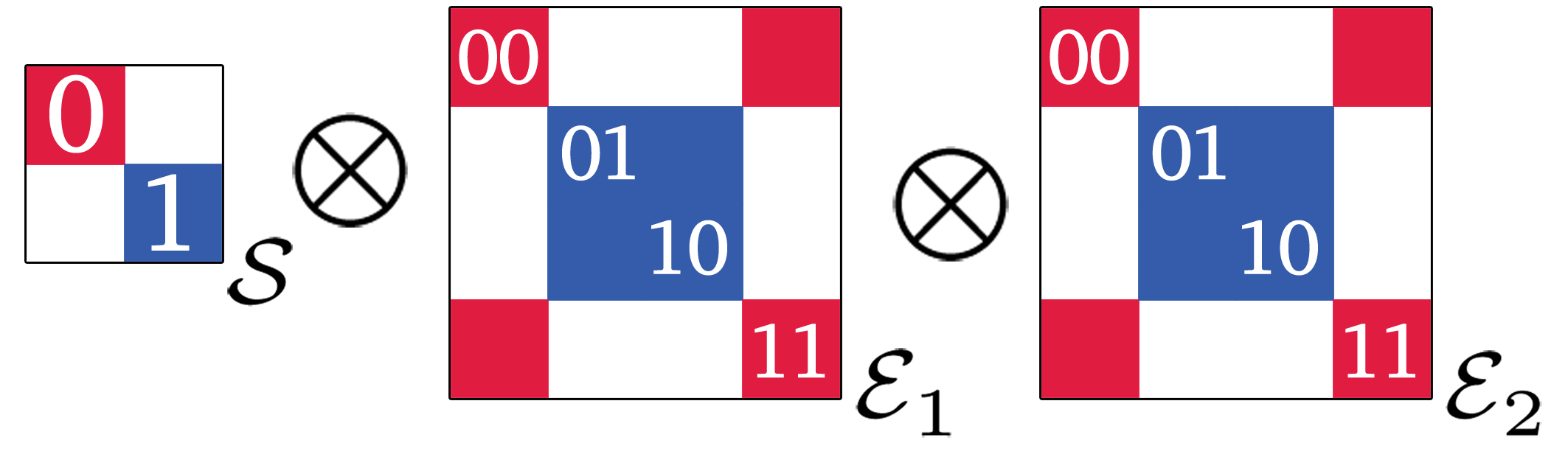}
\par\end{centering}
\caption{Visualisation of our pre-determined objective subspace. The system and environments must be in a statistical mixture of the red states and the blue states. For example, if the system in state $\left|0\right\rangle_{\mathcal{S}}$ then each environment should be in the space spanned by $\left\{ \left|00\right\rangle,\left|11\right\rangle\right\}$. \label{fig:Visualisation-of-the-objective-subspaces}}
\end{figure}

The system is comprised on one photonic polarisation qubit, with fixed basis $\ket{0}_{\mathcal{S}}$ and $\ket{1}_{\mathcal{S}}$ (corresponding to horizontal and vertical polarisation respectively). In general, the dimension of each environment can be larger than the dimension of the system. Here, we consider each environment $\mathcal{E}_{k}$ as being composed of two photons, $\left(\mathcal{E}_{k}^{\left(1\right)},\mathcal{E}_{k}^{\left(2\right)}\right)$. The parity of the environment state corresponds to the two disjoint subspaces that signal strong quantum Darwinism, which we depict in Fig.~\ref{fig:Visualisation-of-the-objective-subspaces}. If the system in state $\ket{0}_{\mathcal{S}}$ then the environment should be in the space spanned by $\left\{ \ket{00},\ket{11}\right\} $; and if the system is in state $\ket{1}_{\mathcal{S}}$ then the environment should be in the space spanned by $\left\{ \ket{01},\ket{10}\right\} $. The environment is comprised of two subenvironments $\mathcal{E}_{1}$ and $\mathcal{E}_{2}$. This allows us to probe non-objectivity for fragments $\mathcal{F}=\mathcal{E}_{1}$ and $\mathcal{F}=\mathcal{E}_{1}\mathcal{E}_{2}$. 

\begin{figure}
\begin{centering}
\includegraphics[width=1\textwidth]{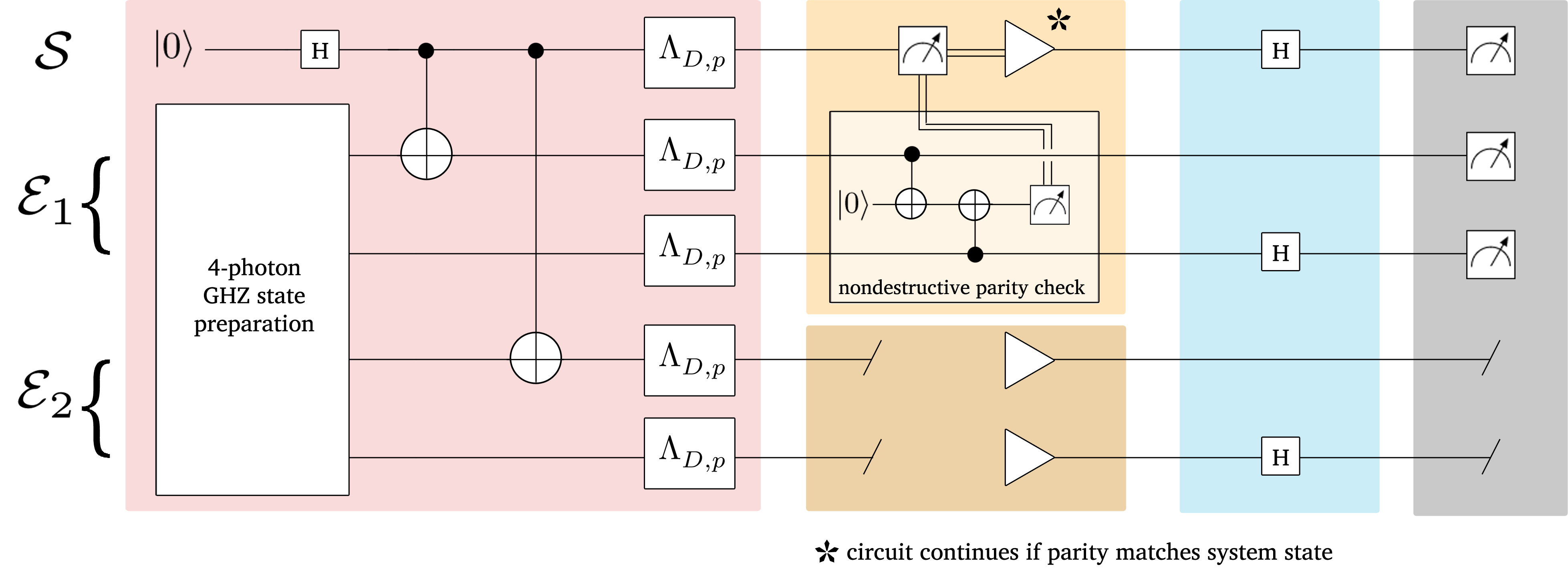}
\par\end{centering}
\caption{\textbf{Circuit for one particular run within the scheme to witness non-objectivity}. Each environment consists of two photons. The system-environment is first prepared into the state given in Eq.~(\ref{eq:initial_state}), including 4-photon GHZ state preparation shown in further detail in Fig.~\ref{fig:GHZ_element}. In this particular run, we have depolarisation channels $\Lambda_{D,p}$ [Eq.~(\ref{eq:depolarisation_channel})] on all photons. The non-fragment component $\mathcal{E}\backslash\mathcal{F}$, here $\mathcal{E}_{2}$, undergoes a point channel, where the photons are discarded and uncorrelated photons are produced. Meanwhile, the system $\mathcal{S}$ is measured and $\mathcal{E}_{1}$ undergoes a nondestructive parity check involving an auxiliary photon. If the parity measurement matches the system measurement result, then the measured system state is recreated, and the system-environment has been projected into the objective subspace. The system-environment undergoes unitary evolution, here under Hadamards $H$ (half wave plates). All measurements are in the computational basis of horizontal/vertical polarisation. If the objective operation results in a null state, then all measurement outcomes are zero.\label{fig:Experiment-circuit}}
\end{figure}

The proposed experimental circuit consists of five overall steps (the case of $\mathcal{F}=\mathcal{E}_{1}$ is shown in Fig.~\ref{fig:Experiment-circuit}):
\begin{enumerate}
\item Preparing the states with some non-objectivity that we wish to witness,
\item Applying a point channel on $\mathcal{E}\backslash\mathcal{F}$,
\item Applying the objectivity operation on $\mathcal{SF}$, or an identity operation, depending on the run,
\item The unitary evolution,
\item And the final measurement.
\end{enumerate}

Our initial system-environment state is
\begin{equation}
\ket{\Psi\left(0\right)}_{\mathcal{SE}_{1}\mathcal{E}_{2}}=\dfrac{1}{2}\ket{0}_{\mathcal{S}}\left(\ket{00,00}_{\mathcal{E}_{1}\mathcal{E}_{2}}+\ket{11,11}_{\mathcal{E}_{1}\mathcal{E}_{2}}\right)+\dfrac{1}{2}\ket{1}_{\mathcal{S}}\left(\ket{10,10}_{\mathcal{E}_{1}\mathcal{E}_{2}}+\ket{01,01}_{\mathcal{E}_{1}\mathcal{E}_{2}}\right),\label{eq:initial_state}
\end{equation}
which has strong Quantum Darwinism when one environment is traced out, and is entangled over the full environment. From this base initial state, we will consider two different operations that reduce the objectivity in the reduced environment state: either mixing the initial state with the maximally noisy state $\id_{\mathcal{S}\mathcal{E}_{1}\mathcal{E}_{2}}/d_{\mathcal{S}}d_{\mathcal{E}_{1}}d_{\mathcal{E}_{2}}$, such that the state at time $t$ is 
\begin{equation}
\rho_{\mathcal{SE}_{1}\mathcal{E}_{2}}\left(t\right)=\left(1-p\right)\ket{\Psi\left(0\right)}\bra{\Psi\left(0\right)}_{\mathcal{SE}_{1}\mathcal{E}_{2}}+p\dfrac{\id_{\mathcal{S}\mathcal{E}_{1}\mathcal{E}_{2}}}{d_{\mathcal{S}}d_{\mathcal{E}_{1}}d_{\mathcal{E}_{2}}},\label{eq:mix_with_maximally_noisy}
\end{equation}
or applying local depolarisation on all photons:
\begin{equation}
\Lambda_{D,p}^{X}\left(\rho_{X}\right)=\left(1-p\right)\rho_{X}+p\dfrac{\id_{X}}{d_{X}}\rho,\qquad X=\mathcal{S},\mathcal{E}_{1}^{\left(1\right)},\mathcal{E}_{1}^{\left(2\right)},\mathcal{E}_{2}^{\left(1\right)},\mathcal{E}_{2}^{\left(2\right)},\label{eq:depolarisation_channel}
\end{equation}
where $\mathcal{E}_{1}^{\left(a\right)}$, $a=1,2$ denotes the two sub photons in environment $\mathcal{E}_{1}$ etc.

The objective operation involves measuring the system and applying a non-destructive parity measurements on each environment in the fragment. If the results match then the system-fragment is objective in our pre-determined subspace. The nondestructive parity measurement on the environment means that we only need to reconstruct the system state after its measurement.

In the unitary phase of the witness, we apply Hadamards $H$ on alternating photons: $U=\left(H\right)_{\mathcal{S}}\otimes\left(\id_{2}\otimes H\right)_{\mathcal{E}_{1}}\otimes\left(\id_{2}\otimes H\right)_{\mathcal{E}_{2}}$.

The final measurements are all in computational basis, i.e. in the horizontal/vertical polarisation basis.

In the following subsection \ref{subsec:Experimental-Components} we include some basic experimental detail on a hypothetical experiment. The numerical simulation of the circuit is given in subsection \ref{subsec:Numerical-Simulations}.

\subsection{Experimental Components\label{subsec:Experimental-Components}}

There are a number of major components in that circuit, which we will go in to with further detail: the large number of controlled-NOT (CNOT) operations which are required for the state preparation and the objectivity operation, the procedure to generate four-photon GHZ states required for the initial state preparation, and the  objectivity operation itself.

\subsubsection{Controlled-NOT Operation}

The experimental procedure uses a number of nondestructive CNOT gates: in the preparation of the initial state in Eq.~(\ref{eq:initial_state}) and for the parity measurement for the objective operation, both seen in Fig.~\ref{fig:Experiment-circuit}. This is also the main limiting factor of the witnessing scheme, and the scalability of the witnessing scheme depends heavily on the success probability and fidelity of the CNOT gates (see subsection \ref{subsec:Comparison-with-State-Tomography} on this scaling).

\begin{figure}
\begin{centering}
\includegraphics[width=1\textwidth]{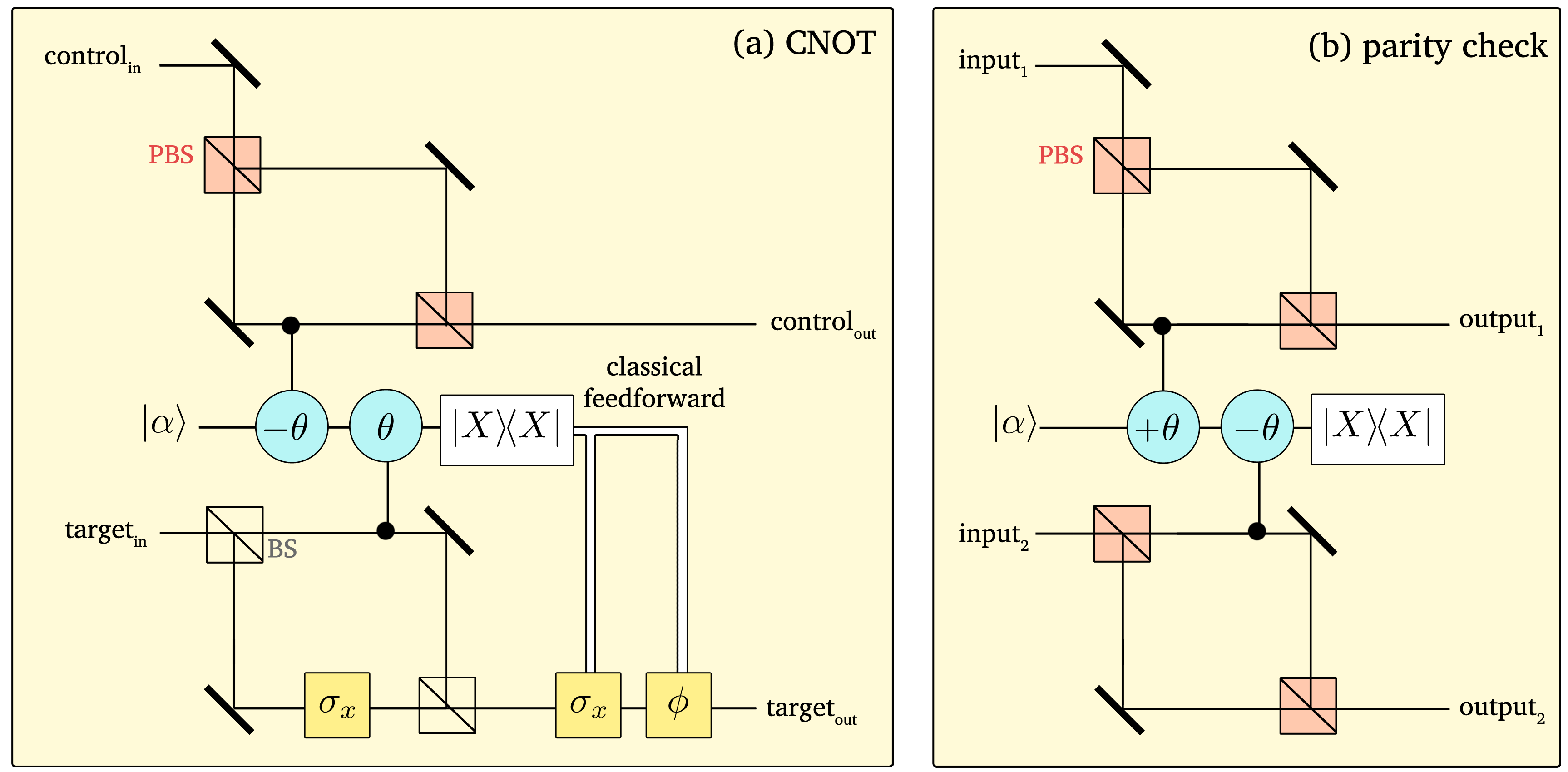}
\par\end{centering}
\caption{(a) Controlled-NOT for two photon polarisation qubits from Ref. \citep{Lin2009}. This uses a cross-Kerr nonlinearity in the controlled-phase gates $\pm\theta$ to boost gate success probability to $1/2$. PBS denotes polarisation beam splitter, and BS denotes beam splitter. An auxiliary coherent state $\left|\alpha\right\rangle $ is used. $\left|X\right\rangle \left\langle X\right|$ is a X homodyne measurement, and depending on its results, a further $\sigma_{x}$ gate and phase $\phi$ gate may be applied to the target photon. 
(b) Nondestructive parity measurement from Ref. \citep{Nemoto2004}. Aside from polarising beam splitters (PBS), it uses cross-Kerr nonlinearities that apply a shift of $+\theta$, and $-\theta$ on the coherent state $\left|\alpha\right\rangle $ if there is photon in the corresponding control modes, followed by a X homodyne measurement $\left|X\right\rangle \left\langle X\right|$ in order to determine the parity $\left\{ \left|00\right\rangle ,\left|11\right\rangle \right\} $ or $\left\{ \left|01\right\rangle ,\left|10\right\rangle \right\} $ of the input photons. \label{fig:CNOT_parity}}
\end{figure}

If we use only linear optical elements, then CNOTs gates are necessary probabilistic with realistic success probabilities of $1/4$ to $1/16$ \citep{Gasparoni2004}. In contrast, if we employ nonlinearities, then we could make deterministic or near-deterministic CNOTs \citep{Nemoto2004,Lin2009} with greater success probabilities. One explicit example is given in Fig.~\ref{fig:CNOT_parity}(a), which uses cross-Kerr nonlinearity to boost success probability up to $1/2$.

\subsubsection{Initial state preparation}

\begin{figure}
\begin{centering}
\includegraphics[width=0.9\textwidth]{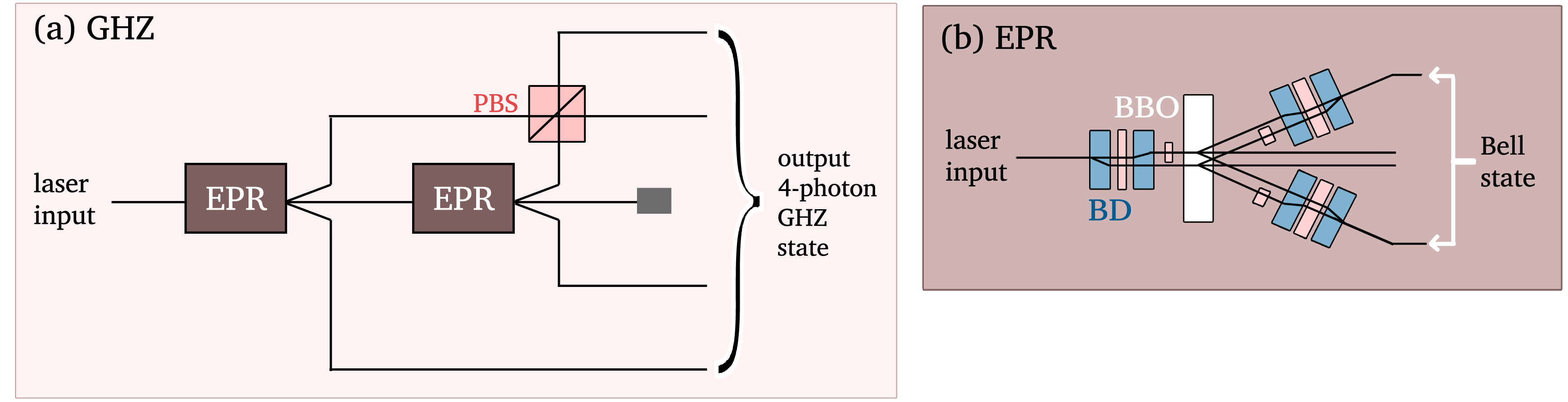}
\par\end{centering}
\caption{\textbf{Initial state preparation}. (a) Four-photon Greenberger-Horne-Zeilinger (GHZ) state preparation: producing two Bell states (via the EPR elements), then combining one photon from each pair at a polarising beam splitter (PBS) to produce a four-photon GHZ state. (b) The EPR element to create Bell states using spontaneous parametric down-conversion: consists of passing laser pulses through betabarium borate (BBO) nonlinear crystals and beam displacers (BDs). Adapted from Ref. \citep{Chen2019}. \label{fig:GHZ_element}}
\end{figure}

The system state can be created with a Hadamard $H$ produced with a half wave plate (HWP) at $\theta=\pi/2$: $\left(\ket{0}+\ket{1}\right)/\sqrt{2}=H\ket{0}$. This is shown in the first left-hand box in Fig.~\ref{fig:Experiment-circuit}. We then need to create a four-photon Greenberger-Horne-Zeilinger (GHZ) state on the environment. This procedure is depicted in Fig.~\ref{fig:GHZ_element}, and proceeds as follows: we would first create Bell states $(\ket{00}+\ket{11})/\sqrt{2}$ on each environment via spontaneous parametric down-conversion (SPDC). One photon from each pair then goes into a beam splitter with path lengths such that they arrive at the same time. Coincident detection in the outputs implies that each both photons are $H$-polarised or $V$-polarised, corresponding to projecting the four photons into the subspace spanned by $\left\{ \ket{00,00},\ket{11,11}\right\} $. After renormalising the state, the four-photon GHZ state is made. The probability of the four-photon GHZ state being made is $1/4$ \citep{Zhao2004}.

Finally, we would apply CNOT operations, with the system as the control, onto the first photons of each of the two environments of the following system-environment state,
\begin{equation}
CNOT_{\mathcal{S}:\mathcal{E}_{1}^{\left(1\right)}}CNOT_{\mathcal{S}:\mathcal{E}_{2}^{\left(2\right)}}\left[\dfrac{1}{\sqrt{2}}\left(\ket{0}+\ket{1}\right)_{\mathcal{S}}\otimes\dfrac{1}{\sqrt{2}}\left(\ket{00,00}+\ket{11,11}\right)_{\mathcal{E}_{1}\mathcal{E}_{2}}\right],
\end{equation}
to produce the initial state given in Eq.~(\ref{eq:initial_state}).

The second component of state preparation is degrading the objectivity in the system-environment. Mixing with the maximally noisy state [Eq.~(\ref{eq:mix_with_maximally_noisy})] can be done simply by creating the initial state with probability $1-p$, and the maximally noisy state with probability $p$. Meanwhile, for the depolarisation channel, one possibility is to use the circuit described in Ref. \citep{Jeong2013}. An alternative procedure is to mix the local state of each photon with the local unpolarised state, i.e. keeping a photon with $1-p$ or replacing it with a completely unpolarised state with probability $p$ (e.g. by passing it through a multimode fibre, or by simply generating a new unpolarised state). By averaging over sufficiently many runs, this would give an effective depolarisation channel.

\subsubsection{Objectivity Operation}

The objectivity operation from Eq.~(\ref{eq:SQD_objective_op}) can be implemented with a system polarisation measurement and nondestructive parity checks on the environment. One possible parity check scheme is shown in Fig.~\ref{fig:Experiment-circuit}, which uses an auxiliary photon and two CNOT operations. Alternatively, a direct parity check scheme could be used, such as the one from \citep{Nemoto2004}, reproduced in Fig.~\ref{fig:CNOT_parity}(b). Either methods employ nonlinear cross-Kerr nonlinearities, which is crucial for increasing the success probability of the operation.

\subsection{Numerical Simulations\label{subsec:Numerical-Simulations}}

In this section, we numerically simulate the circuit and scheme we proposed. The results are given  in Figs. \ref{fig:Numerical-Simulation-SQD} and \ref{fig:Numerical-Simulation-SQD-imperfect}. Fig.~\ref{fig:Numerical-Simulation-SQD} shows the presumes a perfect circuit, while Fig.~\ref{fig:Numerical-Simulation-SQD-imperfect} considers if the state preparation involved controlled-NOT operations
with fidelities of approximately $0.79$. We find that our witness is able to detect non-objectivity in the cases we considered.

\subsubsection{Measurement Operators in the Witness}

\begin{figure}
\begin{centering}
\includegraphics[width=0.9\textwidth]{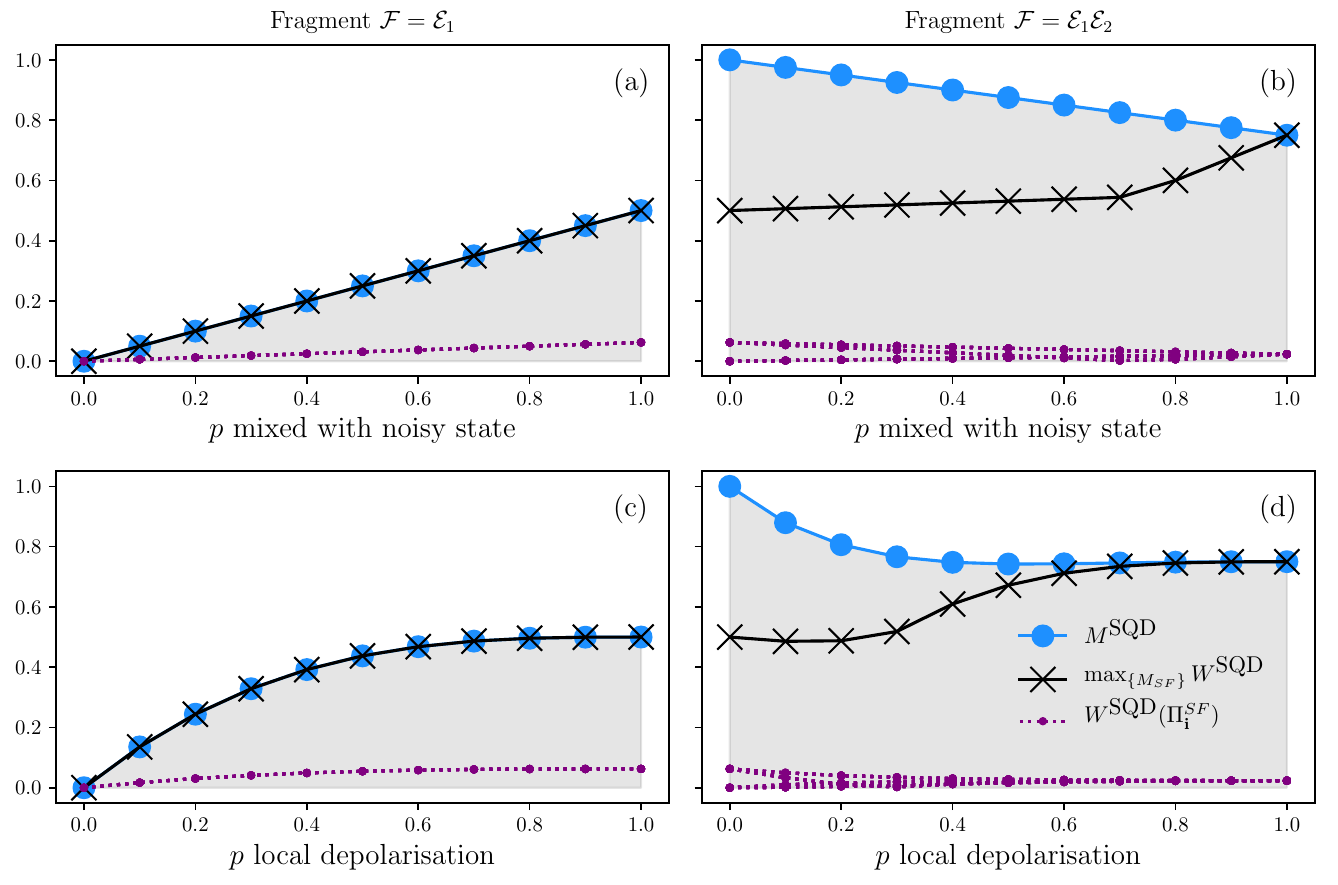}
\par\end{centering}
\caption{\textbf{Numerical Simulation of the non-objectivity witness in the strong quantum Darwinism framework. }Fragments either consisted of the first environment $\mathcal{E}_{1}$ [(a), (c)] or both environments $\mathcal{E}_{1}\mathcal{E}_{2}$ [(b), (d)]. $M^{\text{SQD}}$ refers to the subspace-dependent non-objectivity measure defined in Eq.~(\ref{eq:measureSQD}); $\max_{\left\{ M_{\mathcal{SF}}\right\} }W^{\text{SQD}}$ refers to the maximum value of the witness when the final measurement $M_{\mathcal{SF}}$ is the summation of a subset of computational projective measurement operators [Eq.~(\ref{eq:maxWSQD_sum})]; and the $W^{\text{SQD}}\left(\Pi_{\boldsymbol{i}}^{\mathcal{SF}}\right)$ correspond to the values of the witness with rank-1 measurement $\Pi_{\boldsymbol{i}}^{\mathcal{SF}}$ in the computational basis $\left\{ \left|\boldsymbol{i}\right\rangle \right\} _{\boldsymbol{i}}$. The probability $p$ is the additional noise, either due to mixing with the noisy state Eq.~(\ref{eq:mix_with_maximally_noisy}) or local depolarisation Eq.~(\ref{eq:depolarisation_channel}) on the initial state.\label{fig:Numerical-Simulation-SQD}}
\end{figure}

The final measurements of the system and fragment are always in the computational basis. This reflects our motivation that only particular measurements are possible, or preferable, in a realistic system. For a particular projective rank-1 measurement $\Pi_{\boldsymbol{i}}^{\mathcal{SF}}$ in the computational basis, the non-objectivity witness $W^{SQD}\left(\Pi_{\boldsymbol{i}}^{\mathcal{SF}}\right)$ has a very small value in general, and this can be seen near the bottom of the plots in Figs. \ref{fig:Numerical-Simulation-SQD} and \ref{fig:Numerical-Simulation-SQD-imperfect}. However, we can construct a larger witness without any additional measurements by considering a \emph{collection} of the computational measurement operators $\Pi_{\boldsymbol{i}}^{\mathcal{SF}}$:
\begin{equation}
\max_{\left\{ M_{\mathcal{SF}}\right\} }W^{SQD}=\max_{\left\{ M_{\mathcal{SF}}\right\} }\left|\sum_{\Pi_{\boldsymbol{i}}^{\mathcal{SF}}\in\left\{ M_{\mathcal{SF}}\right\} }\tr\left[\Pi_{\boldsymbol{i}}^{\mathcal{SF}}\left\{ \rho_{\mathcal{SE}}\left(t+\tau\right)-\rho_{\mathcal{SE}}^{\prime}\left(t+\tau\right)\right\} \right]\right|.
\end{equation}
As the term within the absolute magnitude, $\tr\left[\Pi_{\boldsymbol{i}}^{\mathcal{SF}}\left\{ \rho_{\mathcal{SE}}\left(t+\tau\right)-\rho_{\mathcal{SE}}^{\prime}\left(t+\tau\right)\right\} \right]$,
can be either positive or negative, this leads to:
\begin{eqnarray}
\max_{\left\{ M_{\mathcal{SF}}\right\} }W^{\text{SQD}} & =&\max\biggl\{\biggl|\sum_{\Pi_{\boldsymbol{i}}^{\mathcal{SF}}\text{ s.t. }\tr\left[\cdots\right]\geq0}\tr\left[\Pi_{\boldsymbol{i}}^{\mathcal{SF}}\left\{ \rho_{\mathcal{SE}}\left(t+\tau\right)-\rho_{\mathcal{SE}}^{\prime}\left(t+\tau\right)\right\} \right]\biggr|,\label{eq:maxWSQD_sum}\\
 & &\phantom{===\max}\biggl|\sum_{\Pi_{\boldsymbol{i}}^{\mathcal{SF}}\text{ s.t. }\tr\left[\cdots\right]<0}\tr\left[\Pi_{\boldsymbol{i}}^{\mathcal{SF}}\left\{ \rho_{\mathcal{SE}}\left(t+\tau\right)-\rho_{\mathcal{SE}}^{\prime}\left(t+\tau\right)\right\} \right]\biggr|\biggr\}.\nonumber 
\end{eqnarray}
The value of $\max_{\left\{ M_{\mathcal{SF}}\right\} }W^{\text{SQD}}$, as can be seen in Fig.~\ref{fig:Numerical-Simulation-SQD} can provide a good, sometimes\emph{ tight}, lower bound to the true value of the measure $M^{\text{SQD}}$. The bound is less tight when the fragment consists of the full environment, $\mathcal{F}=\mathcal{E}_{1}\mathcal{E}_{2}$: these situations correspond to greater quantum correlations between the system and fragment. This is expected, as the unitary evolution is local, and the measurements are also local, and so are unable to capture the full quantumness component of the non-objectivity. Thus, one possible improvement would be to introduce an entangling or global unitary evolution. Despite this, the witness \emph{is} successful in detecting non-objectivity.

\subsubsection{Simulation with operation fidelities and noise \label{subsec:noisy_numerical_simulation}}

\begin{figure}
\begin{centering}
\includegraphics[width=0.5\textwidth]{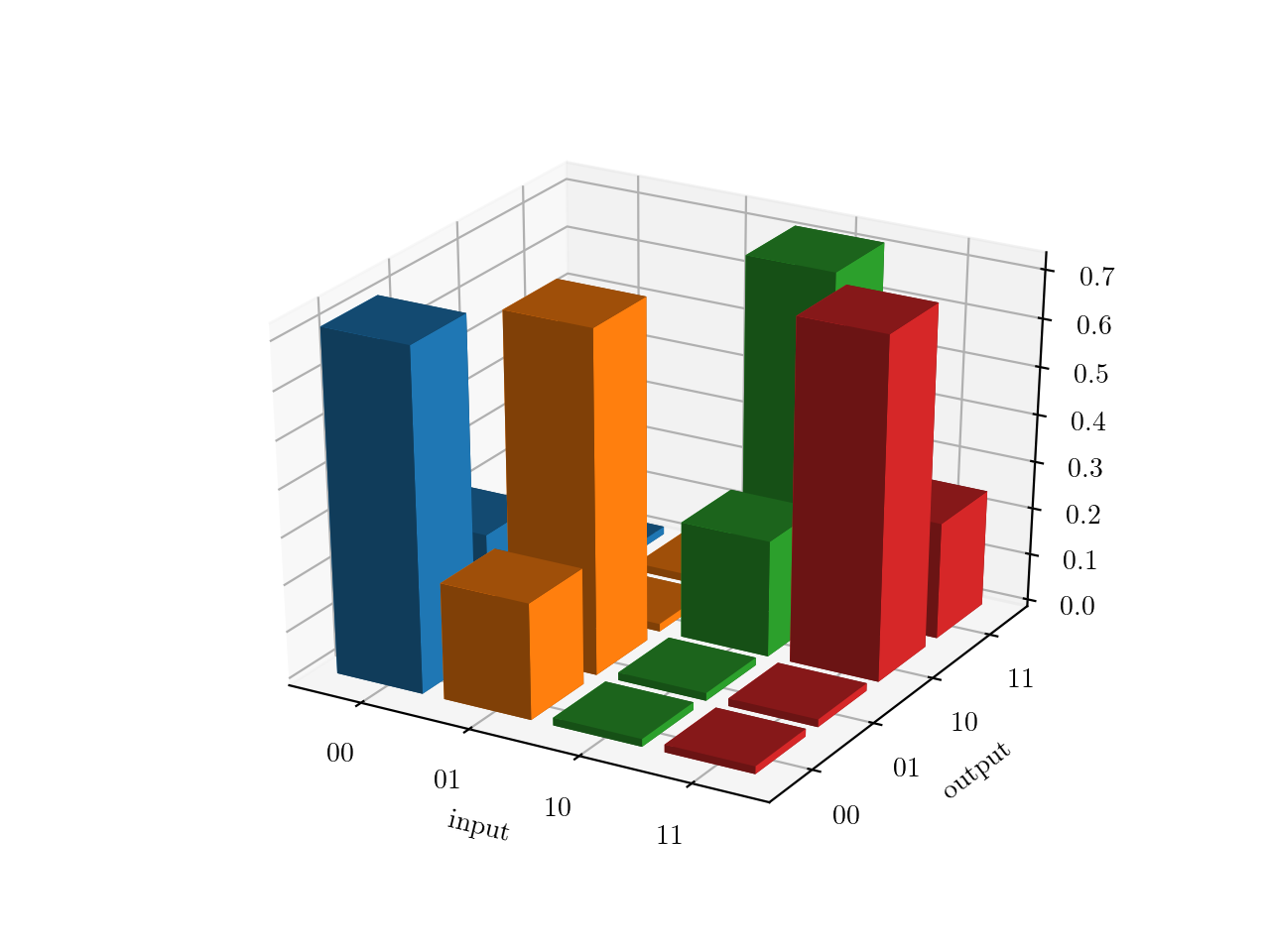}
\par\end{centering}
\caption{\textbf{Simulated CNOT logic table with average gate fidelity of $0.79$} used for the numerical simulation in Fig.~\ref{fig:Numerical-Simulation-SQD-imperfect}. \label{fig:numerical_CNOT}}
\end{figure}

In a realistic experiment, gate operations are not perfect. In particular, we have employed various CNOT operations which can have fidelity $F<1$. To simulate such results, we consider a CNOT gate that behaves imperfectly, with the given logic table in Fig. \ref{fig:numerical_CNOT}, whose probabilities are approximations of experimental CNOT logic tables \citep{Zhao2005,Bao2007,Zeuner2018}. The average simulated gate fidelity is $\bar{F}\approx0.79$.

\begin{figure}
\begin{centering}
\includegraphics[width=0.9\textwidth]{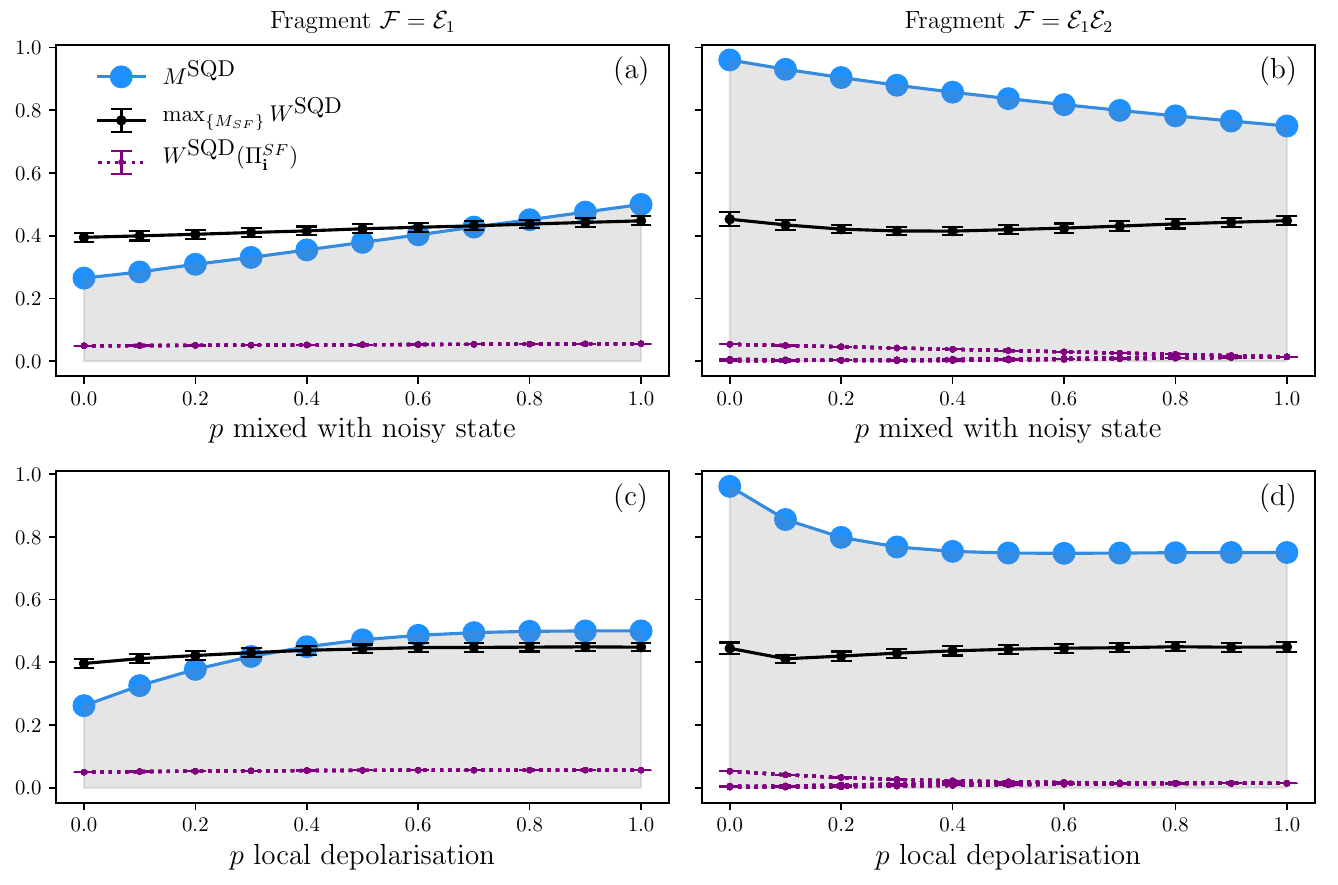}
\par\end{centering}
\caption{\textbf{Numerical Simulation of the non-objectivity witness in strong quantum Darwinism framework with imperfect state preparation and gates}. Details are given in Sec.~\ref{subsec:noisy_numerical_simulation}. Briefly, we have 500 initial states are randomly distributed close to the wanted initial state Eq.~(\ref{eq:initial_state}), 1000 copies of each random initial state, and stochastic measurement containing Poisson noise distributed with mean photon number $N_p=100$. Monte Carlo averaging occurred over measurements and states, and the witness is averaged over all random initial states.} $M^{\text{SQD}}$ refers to the subspace-dependent measure defined in Eq.~(\ref{eq:measureSQD}), given for the (imperfect) state at time ``$t$'' of Fig.~\ref{fig:Protocol-for-a-witness} (c.f. $M^{\text{SQD}}$ in Fig.~\ref{fig:Numerical-Simulation-SQD}); $\max_{\left\{ M_{\mathcal{SF}}\right\} }W^{\text{SQD}}$ refers to the maximum value of the witness when the final measurement $M_{\mathcal{SF}}$ is the summation of a subset of computational projective measurement operators (in Eq.~(\ref{eq:maxWSQD_sum})); and the $W^{\text{SQD}}\left(\Pi_{\boldsymbol{i}}^{\mathcal{SF}}\right)$ correspond to the values of the witness with rank-1 measurement $\Pi_{\boldsymbol{i}}^{\mathcal{SF}}$ in the computational basis $\left\{ \left|\boldsymbol{i}\right\rangle \right\} _{\boldsymbol{i}}$. The probability $p$ is the additional noise, either due to mixing with the noisy state  Eq.~(\ref{eq:mix_with_maximally_noisy}) or local depolarisation Eq.~(\ref{eq:depolarisation_channel}).
\label{fig:Numerical-Simulation-SQD-imperfect}
\end{figure}

Fig.~\ref{fig:Numerical-Simulation-SQD-imperfect} gives the results of stochastic numerical simulation. Along with the noisy CNOT gates described above, the simulation is performed thus:

\begin{enumerate}
    \item Choose $N_{\rho}=500$ random initial states $\rho_i$ distributed close to the ideal state given in Eq.~(\ref{eq:initial_state}).
    \item For each $\rho_i$, simulate $N_c=1000$ stochastic runs of the circuit.
    \item For each run, simulate measurement with total photon number picked from a Poisson distribution with mean photon number $N_p=100$.
    \item For each $\rho_i$, average over all measurement outcomes and calculate the optimal witness, $\max_{\{M_{SF}\}}W^{\text{SQD}}$
    \item Calculate the average optimal witness over all random initial states.
\end{enumerate}

Thus, the simulation results in Fig.~\ref{fig:Numerical-Simulation-SQD-imperfect} have three primary noise sources: the initial imperfect preparation, the nonzero fidelity into the CNOT gates in the objective operation, and measurement noise.

We can see in Fig.~\ref{fig:Numerical-Simulation-SQD-imperfect} that the simulated experimental witness can overshoot the true measure values, especially when there is less non-objectivity [see ]Fig.~\ref{fig:Numerical-Simulation-SQD-imperfect}(a), (c)]. Therefore, experimentally, the witness should pass nonzero lower bound before non-objectivity can be declared. As the CNOT gate fidelities increase, the optimal witnesses of Fig.~\ref{fig:Numerical-Simulation-SQD-imperfect}(a), (c) approach the true value of the measure (not shown here). Also note that the measure is never zero at $p=0$ unlike in exact case of Fig.~\ref{fig:Numerical-Simulation-SQD}: due to the imperfect system-environment state preparation, the initial states are   non-objective.

Without an exact experimental set up, it is unfeasible to model real and detailed error mechanisms. Nonetheless, the results of Fig.~\ref{fig:Numerical-Simulation-SQD-imperfect} provide an initial demonstration of a potential future experiment, showing that the scheme could realistically detect non-objectivity.

\subsection{Comparison with quantum state tomography\label{subsec:Comparison-with-State-Tomography}}

Let us make a back-of-the-envelope comparison between quantum state tomography versus our witness based on the number of trials required in either scheme to produce statistics with similar confidence. We will not count the preparation component for the state at time $\rho_{\mathcal{SE}}\left(t\right)$ as this is required regardless of technique. Rather, we want to compare the number of required measurements in the multiple different bases needed for state tomography, versus the witness which involves two different sets of runs split into (1) point channel, unitary evolution (Hadamards), and computational measurements in the horizontal/vertical polarisation basis and (2) point channel, objectivity operation, unitary evolution and the computational measurements.

Let us suppose there are $1+2M$ photons in $\mathcal{SF}$ (one system photon, and $M$ environments with $2$ photons in each environment): 
For the state tomography, the photons in $\mathcal{SF}$ need to be measured in three different bases (local observables), in all the combinations (i.e. with polarisation analysis in horizontal/vertical, left/right, and diagonal/anti-diagonal polarisation measurements). With $1+2M$ photons to measure, there are $3^{1+2M}$ basis combinations. Let $C$ be the number of counts in one basis set required for sufficient statistics. Then quantum state tomography would naively require $C\cdot3^{1+2M}$ runs in total.

Meanwhile, for the witness, we assume that the point channel and unitary evolutions are deterministic (such as the Hadamards we have chosen). And since we fix the final end basis, the first set requires $C$ runs. The second set of runs requires the objectivity operation. Let $p_{CNOT}$ be the probability of the CNOT operation succeeding. Two CNOT operations are required per environment, therefore the probability of a successful parity check on one environment has probability $p_{CNOT}^{2}.$ There are $M$ environments, therefore the probability of all the parity checks successfully occurring is $\left(p_{CNOT}^{2}\right)^{M}$. In order for there to be a total of $C$ successful runs, there must have been at least $C\left(1/p_{CNOT}\right)^{2M}$ copies of the state. Therefore, the witness scheme would require $C+C\left(1/p_{CNOT}\right)^{2M}$ runs in total.

If the CNOT operation can be implemented with success probability of $p_{CNOT}\gtrsim1/3$, then the witness scheme outperforms quantum state tomography. If we introduce in the fidelities for the CNOT operation, then could roughly replace $p_{CNOT}\rightarrow p_{CNOT}\cdot F_{CNOT}$ where $F_{CNOT}$ is the fidelity. If $F_{CNOT}=0.79$, then a minimum success probability of $p_{CNOT}\geq1/3F_{CNOT}\approx0.42$ is required. In previous literature, for example Ref. \citep{Lin2009}, a theoretical $p_{CNOT}=1/2$ is possible, using nonlinear elements. Although a realistic device suffers from decoherence effects \cite{He2012,Fan2013}, there has been much work in analysing it and reducing it in order to increase success probabilities and fidelities \cite{Heo2017,Heo2018,Hong2019}. This shows that with sufficiently good controlled-NOT operations with reasonable success probability, the witnessing scheme presented here can scale better as more environments are added.

\section{Discussion\label{sec:Discussion}}

We presented a subspace-dependent witness of non-objectivity in the framework of strong quantum Darwinism. We defined an \emph{objectivity operation} that projects the  system-environment into a preferred objective subspace. The witness detects non-objectivity by comparing the evolution of the system-environment with and without the objectivity operation. We used a point channel to ensure that there is no correlations between the system-fragment and the rest of the environment that might accidentally trigger the witness. We showed that this witness scheme has the potential to scale better than quantum state tomography. Hence, our witnessing scheme and proposal opens up a pathway to experimental verification of Quantum Darwinism in increasingly large system-environments.

Our witness scheme relies on projective measurements. Therefore, photonic systems are ideal, and are our chosen system in our experimental proposal. In contrast, the ensemble measurements in spin and magnetic resonance systems are non-projective \cite{Vandersypen2005, Laflamme2002}. Projective measurement results can be simulated by measuring multiple observables or with other schemes \cite{Lee2006}.

Secondly, our witness scheme relies on non-invasive/non-demolition CNOT and parity check operations on photonic qubits to be realised via cross-Kerr nonlinearities. In general, such nonlinearities are difficult to implement mostly due to the extremely weak interactions between photons. However, using strong coherent pulses can reduce the required nonlinearity to orders $\theta=10^{-5}$ \cite{Munro2005} to $\theta = 3\times10^{-2}$ \cite{Nemoto2004} such that \emph{strong} cross-Kerr linearities are not necessary. A few years ago, Ref.~\cite{Venkataraman2012} was able to achieve nonlinearities up to $\theta  = 10^{-2}$ using rubidium atoms confined in photonic bandgap fibres, which therefore means that our scheme's use of non-invasive operations is within experimental reach. It is also possible to achieve large phase shifts in photons without using Kerr nonlinearities \citep{Tiarks2016, Ramelow2019, Tiarks2018,Hacker2016}.

An interesting aspect of our approach is that when all the dimensions of the environments are equal to the dimension of the system, then strong quantum Darwinism and spectrum broadcast structure collapse to invariant spectrum structure (\textbf{ISBS}) (see Appendix \ref{app:SQD=ISBS}). Such states have the following form:
\begin{equation}
\rho_{\mathcal{S}\mathcal{E}_{1}\cdots\mathcal{E}_{N}}=\sum_{i}p_{i}\ket{i}\bra{i}_{\mathcal{S}}\otimes\ket{i}\bra{i}_{\mathcal{\mathcal{E}}_{1}}\otimes\cdots\otimes\ket{i}\bra{i}_{\mathcal{\mathcal{E}}_{N}},\label{eq:ISBS-discussion}
\end{equation}
for some local diagonal basis $\left\{ \ket{i}_{\mathcal{E}_{k}}\right\} _{i}$ for the various environments. In Appendix \ref{app:Invariant-Spectrum-Broadcast}, we consider the basis-dependent witness of non-objectivity in this framework. In this case, controlled-NOT operations are not required for the objectivity operation, thus greatly simplifying a hypothetical experimental setup.

As for witnessing spectrum broadcast structure: Casting back to the discussion in subsection \ref{subsec:Subspace-dependent-Strong-Quantum-Darwinism}, our scheme would require us to chose a preferred subspace such that each conditional state $\rho_{\mathcal{E}_{1}|i}\otimes\cdots\otimes\rho_{\mathcal{E}_{N}|i}\in\mathcal{H}_{\mathcal{E}_{1}\cdots\mathcal{E}_{N}|i}$
of the broadcast state, 
\begin{equation}
\rho_{\mathcal{S}\mathcal{E}_{1}\cdots\mathcal{E}_{N}}=\sum_{i}p_{i}\ket{i}\bra{i}_{\mathcal{S}}\otimes\rho_{\mathcal{E}_{1}|i}\otimes\cdots\otimes\rho_{\mathcal{E}_{N}|i},
\end{equation}
lies in disjoint subspaces $\mathcal{H}_{\mathcal{E}_{1}\cdots\mathcal{E}_{N}|i}=\mathcal{H}_{\mathcal{E}_{1}|i}\otimes\mathcal{H}_{\mathcal{E}_{2}|i}\otimes\cdots\otimes\mathcal{H}_{\mathcal{E}_{F}|i}$. We can define the projectors $\Pi_{\mathcal{E}_{k}|i}$ into $\mathcal{H}_{\mathcal{E}_{k}|i}$. In strong quantum Darwinism, the environment projector onto each conditional state is simply $\Pi_{\mathcal{F}|i}=\Pi_{\mathcal{E}_{1}|i}\otimes\cdots\otimes\Pi_{\mathcal{E}_{F}|i}$, which preserves some correlations between the different environments. However, that is not allowed in spectrum broadcast structure. A spectrum broadcast structure witness in this style would require environment projectors that \emph{also} break correlations conditional on $i$, which in turn requires discarding the current state and preparing a new one.  The local conditional environment states $\Pi_{\mathcal{E}_{k}|i}\rho_{\mathcal{E}_{k}}\Pi_{\mathcal{E}_{k}|i}$ need to be known, and re-prepared exactly to ensure there are no extraneous correlations. This hypothetical SBS objectivity operation is a much more intensive unwieldy procedure, making spectrum broadcast structure unsuitable for this particular scheme.

It is worth-noting that non-objectivity in a system within the framework of strong quantum Darwinism can arise from two sources: the existence of quantum correlations, or the lack of perfect classical correlations. Our witnessing scheme does not distinguish between the two, instead giving a single measure that captures non-objectivity in its own right. If the source of the non-objectivity is required, we suggest using an extra discord witness \citep{Gessner2013}. This leads to an alternative two-part witness of non-objectivity: a discord witness, followed by some kind of characterisation of the classical information. If only a binary result is required (i.e. whether or not there \emph{is} non-objectivity), then the two-part protocol can terminate the moment the discord is witnessed.

An interesting theoretical extension of this work is a resource theory of non-objectivity, of which our paper provides the first steps. In general, the free states would be the objective states, and the free operations defined as those that transform objective states into objective states \cite{Chitambar2019}. If we apply convex restrictions to the set of objective states---such as the subspace-dependent Quantum Darwinism we presented---then we can apply the prescriptions known for generic convex resource theories, which have known measures and tasks in which those resourceful states are useful \cite{Takagi2019,Chitambar2019}.

The resolution to the quantum-to-classical transition remains elusive. By choosing a naturally preferred basis in which objectivity may arise, we have presented the first witness of non-objectivity, analogous to the witnesses of established non-classical effects like quantum coherence, discord, and entanglement. With gate operations of sufficiently high fidelity and probability of success, the witness scheme scales better than full state tomography. Therefore, our work opens up further experimental development and testing of quantum Darwinism in larger and increasingly realistic scenarios, and thus, composes a step towards understanding the nature of the quantum-to-classical transition.

\subsubsection*{Acknowledgements}

This work was supported by the Engineering and Physical Sciences Research Council {[}grant number EP/L015242/1{]}. We thank J. Lundeen for helpful discussions on the experimental challenges of implementing Kerr-like non-linearities.

\appendix

\section{Strong quantum Darwinism reduces to invariant spectrum broadcast structure when all subsystem dimensions are equal\label{app:SQD=ISBS}}

Invariant spectrum broadcast structure (\textbf{ISBS}) is a particular subcase of spectrum broadcast structure, where the conditional environments states are also pure:
\begin{equation}
\rho_{\mathcal{S}\mathcal{E}_{1}\cdots\mathcal{E}_{N}}=\sum_{i}p_{i}\ket{i}\bra{i}_{\mathcal{S}}\otimes\ket{i}\bra{i}_{\mathcal{\mathcal{E}}_{1}}\otimes\cdots\otimes\ket{i}\bra{i}_{\mathcal{\mathcal{E}}_{N}},\label{eq:ISBS}
\end{equation}
for some local diagonal basis $\left\{ \ket{i}_{\mathcal{E}_{k}}\right\} _{i}$ for the various subenvironments. In this appendix, we will show that strong quantum Darwinism reduces to invariant spectrum broadcast structure when all the dimensions are the same, $d_{\mathcal{S}}=d_{\mathcal{E}_{k}}=d$. In doing so, spectrum broadcast structure also reduces to invariant spectrum broadcast structure.

Strong quantum Darwinism has bipartite spectrum broadcast structure. So first consider 
\begin{equation}
\rho_{\mathcal{SE}_{k}}  =\sum_{i}p_{i}\ket{i}\bra{i}\otimes\rho_{\mathcal{E}_{k}|i},\qquad\rho_{\mathcal{E}_{k}|i}\rho_{\mathcal{E}_{k}|j}=0\quad\forall i\neq j,
\end{equation}
for a single environment $k\in\left\{ 1,\ldots,F\right\} $. The conditional states are orthogonal, where $\rho_{\mathcal{E}_{k}|i}\rho_{\mathcal{E}_{k}|j}=0$. In a space of dimension $d_{\mathcal{E}_{k}}$, all sets of mutually orthogonal states contain at most $d_{\mathcal{E}_{k}}$ state vectors $\left\{ \ket{\psi_{i}}\right\} _{i}$. Since there must be $d_{\mathcal{S}}=d_{\mathcal{E}_{k}}=d$ different states $\rho_{\mathcal{E}_{k}|i}$, this implies that they must be equivalent to one of those sets of mutually orthogonal states, which means that they are pure, i.e. $\rho_{\mathcal{E}_{k}|i}=\ket{\psi_{i}}\bra{\psi_{i}}_{\mathcal{E}_{k}}$. So this shows that locally, the state has invariant spectrum broadcast structure:
\begin{equation}
\rho_{\mathcal{SE}_{k}}  =\sum_{i}p_{i}\ket{i}\bra{i}\otimes\ket{\psi_{i}}\bra{\psi_{i}}_{\mathcal{E}_{k}},
\end{equation}
which is true for every $k\in\left\{ 1,\ldots,F\right\} $.
Now consider the combined state of the system-fragment:
\begin{equation}
\rho_{\mathcal{SF}}  =\sum_{i}p_{i}\ket{i}\bra{i}\otimes\rho_{\mathcal{F}|i}\qquad\rho_{\mathcal{F}|i}\rho_{\mathcal{F}|j}=0\quad\forall i\neq j.
\end{equation}
We have that $\tr_{\mathcal{F}\backslash\mathcal{E}_{k}}\left[\rho_{\mathcal{F}|i}\right]=\rho_{\mathcal{E}_{k}|i}=\ket{\psi_{i}}\bra{\psi_{i}}_{\mathcal{E}_{k}}$. Since the reduced state is pure, $\rho_{\mathcal{E}_{k}|i}$ is not correlated with the other subenvironments in $\rho_{\mathcal{F}|i}$. So along this bipartition, the state is product:
\begin{equation}
\rho_{\mathcal{F}|i}=\ket{\psi_{i}}\bra{\psi_{i}}_{\mathcal{E}_{k}}\otimes\rho_{\mathcal{F}\backslash\mathcal{E}_{k}|i}.
\end{equation}
Repeating this procedure for all the other $\mathcal{E}_{k}$ in $\mathcal{F}$, and one finds that 
\begin{equation}
\rho_{\mathcal{F}|i}=\ket{\psi_{i}}\bra{\psi_{i}}_{\mathcal{E}_{1}}\otimes\cdots\otimes\ket{\psi_{i}}\bra{\psi_{i}}_{\mathcal{E}_{F}},
\end{equation}
i.e. the state has invariant spectrum broadcast structure.

\section{Invariant Spectrum Broadcast Structure Witnessing Scheme\label{app:Invariant-Spectrum-Broadcast}}

Invariant spectrum broadcast structure has a simple structure, describing the situation where a state is objective from the perspective of every subsystem. Analogous to how we fixed the subspaces in strong quantum Darwinism, we fix the basis in ISBS corresponding to a preferred measurement basis. Basis-dependent invariant spectrum broadcast structure has many similarities with incoherent states. However, the set of basis-dependent ISBS states is strictly smaller than the set of incoherent states.

In the pre-determined basis $\{\ket{i}\}_i$, we can define the objectivity operation, 
\begin{equation}
\Gamma_{\mathcal{SF}}^{\text{ISBS}}\left(\rho_{\mathcal{SE}}\right)=\sum_{i}\left( \ket{i\cdots i}\bra{i\cdots i}_{\mathcal{S}\mathcal{F}}\otimes\id_{\mathcal{E}\backslash\mathcal{F}}\right)\,\rho_{\mathcal{SE}}\,\left(\ket{i\cdots i}\bra{i\cdots i}_{\mathcal{S}\mathcal{F}}\otimes\id_{\mathcal{E}\backslash\mathcal{F}}\right).\label{eq:objectivityoperation_ISBS}
\end{equation}
This leads to the corresponding basis-dependent measure of non-objectivity:
\begin{equation}
M^{\text{ISBS}}\left(\rho_{\mathcal{SF}}\left(t\right)\right)=\left\Vert \rho_{\mathcal{SF}}\left(t\right)-\Gamma_{\mathcal{SF}}^{\text{ISBS}}\left(\rho_{\mathcal{SF}}\left(t\right)\right)\right\Vert _{1}.\label{eq:measureISBS}
\end{equation}

The overall witness scheme is identical to that of the strong quantum Darwinism witness scheme in Sec. \ref{sec:Witnessing-non-objectivity}. The corresponding non-objectivity witness in this framework is: 
\begin{equation}
W^{\text{ISBS}}\left(M_{\mathcal{SE}}\right)  =\left|\tr\left[M_{\mathcal{SE}}\mathcal{U}_{\tau}\left[\left\{ \rho_{\mathcal{SF}}\left(t\right)-\Gamma_{\mathcal{SF}}^{\text{ISBS}}\left(\rho_{\mathcal{SF}}\left(t\right)\right)\right\} \otimes\rho_{\mathcal{E}\backslash\mathcal{F}}^{\text{new}}\right]\right]\right|,
\end{equation}
where $M_{\mathcal{SE}}$ is a measurement operator and $\mathcal{U}_\tau$ is some unitary evolution.

\subsection{Quantum Photonic Simulation Proposal}

\begin{figure}
\begin{centering}
\includegraphics[width=0.9\textwidth]{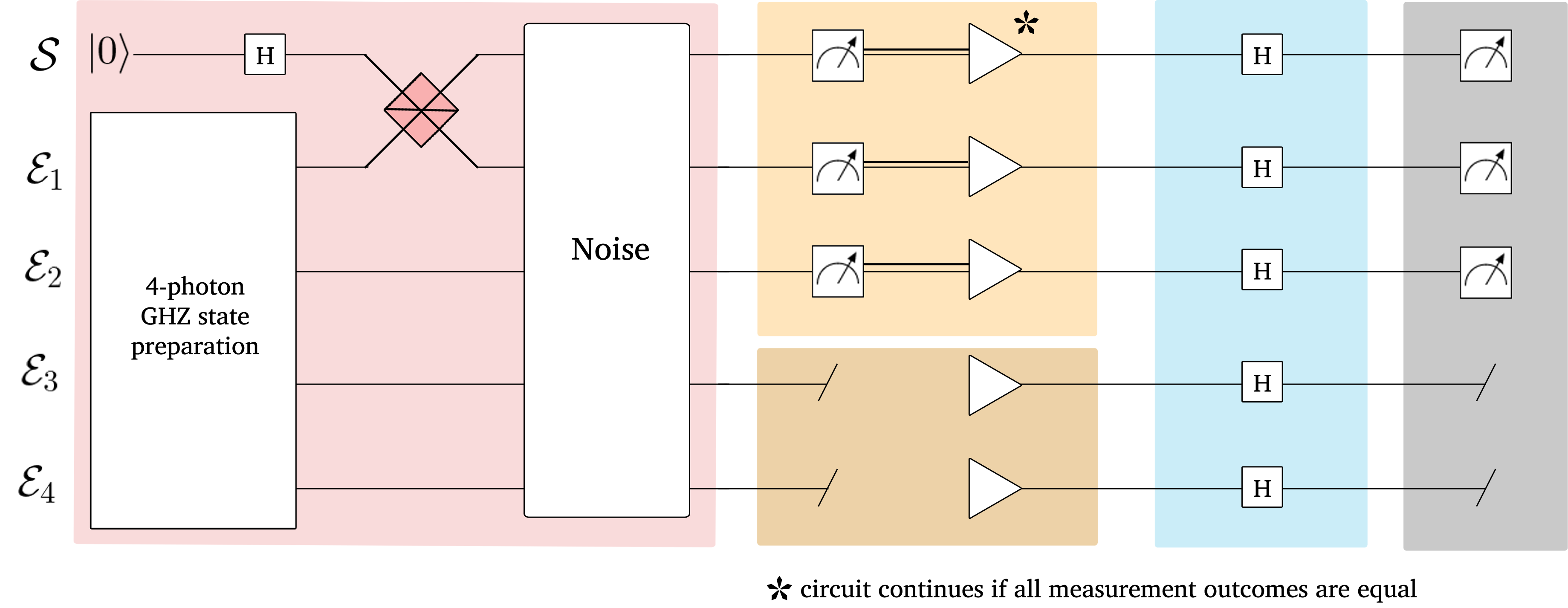}
\par\end{centering}
\caption{\textbf{Circuit for one particular run within the scheme to witness non-objectivity in the invariant spectrum broadcast structure framework}. The system and each environment consists of one a photon polarisation qubit. The system-environment are first prepared in a five-photon GHZ state (starting from the four-photon GHZ from Fig.~\ref{fig:GHZ_element}). In this particular run, we have some noise channel that acts on the system-environment state (Eq.~(\ref{eq:mix_with_maximally_noisy})). The non-fragment component $\mathcal{E}\backslash\mathcal{F}$, here environments  $\mathcal{E}_3$ and $\mathcal{E}_4$, undergo a point channel. Meanwhile, the system and environments $\mathcal{E}_1$ and $\mathcal{E}_2$ are measured. If all measurements results match, corresponding to a successful objective projection, then that state is recreated and sent down the rest of the circuit. The system-environment undergoes unitary evolution, here under Hadamard gates $H$. All measurements are in the computational basis of horizontal/vertical polarisation. \label{fig:Experiment-circuit-ISBS}}
\end{figure}

We take the system as one photon, and consider four environments each comprised of one photon each. The main departure from the scheme with strong quantum Darwinism is the initial state, and the objectivity operation, shown in Fig.~\ref{fig:Experiment-circuit-ISBS} (compare with Fig.~\ref{fig:Experiment-circuit}).

Here, we have initial five-photon GHZ state $\left(\ket{00000}+\ket{11111}\right)/\sqrt{2}$ on the system and environments. This can be created from the extension of the four-photon procedure \citep{Zhao2004}: after creating the four-photon GHZ state for the environment, one prepares the system state in $\left(\ket{0}+\ket{1}\right)/\sqrt{2}$. Then, arrange the path-lengths such that the system photon and the first environment photon arrive at a polarising beam splitter the same time. Similarly, coincident detection in the outputs implies that each both photons are $H$-polarised or $V$-polarised, and after renormalising the state, the five-photon GHZ state is made \citep{Zhao2004}. This is shown on the left-hand-side in Fig.~\ref{fig:Experiment-circuit-ISBS}.

In invariant spectrum broadcast structure, the objectivity operation is now very simple: correlated measurement in the horizontal/vertical polarisation basis in the system and fragment photons. If all measurements on the system and fragment photons give the same outcome, then the system-fragment state is objective, and that state can be recreated and sent down to the rest of the circuit.

\begin{figure}
\begin{centering}
\includegraphics[width=1\textwidth]{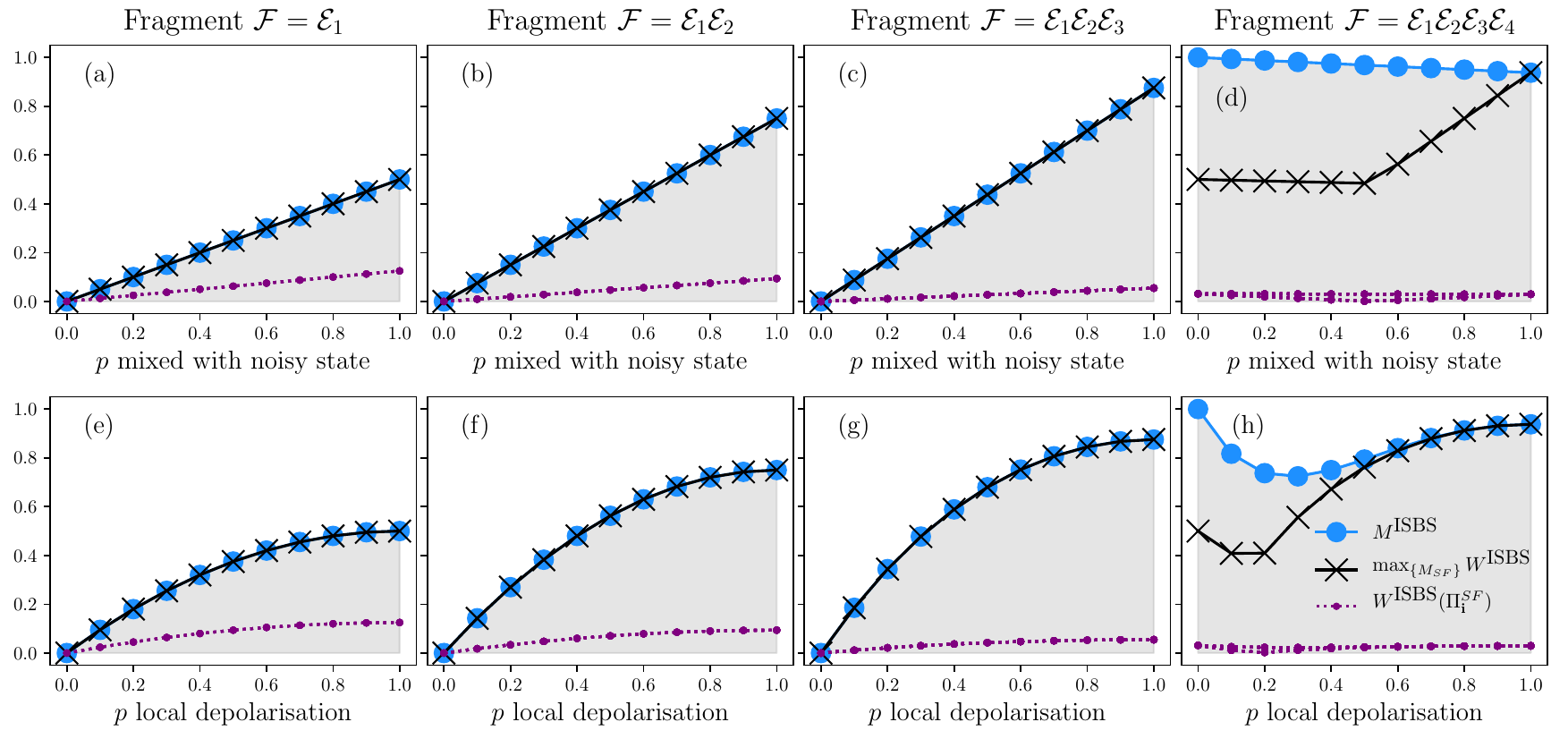}
\par\end{centering}
\caption{\textbf{Numerical Simulation of the non-objectivity witness in the invariant spectrum broadcast structure framework. }Fragments consist of $\left\{ \mathcal{E}_{1}\right\} $, $\left\{ \mathcal{E}_{1},\mathcal{E}_{2}\right\} $, $\left\{ \mathcal{E}_{1},\mathcal{E}_{2},\mathcal{E}_{3}\right\}$ and $\left\{ \mathcal{E}_{1},\mathcal{E}_{2},\mathcal{E}_{3},\mathcal{E}_{4}\right\} $ from left to right. $M^{\text{ISBS}}$ refers to the basis-dependent measure defined in Eq.~(\ref{eq:measureISBS}); $\max_{\left\{ M_{\mathcal{SF}}\right\} }W^{\text{ISBS}}$ refers to the maximum value of the witness when the final measurement $M_{\mathcal{SF}}$ is the summation of a subset of computational projective measurement operators (the ISBS version of Eq.~(\ref{eq:maxWSQD_sum})); and the $W^{\text{ISBS}}\left(\Pi_{\boldsymbol{i}}^{\mathcal{SF}}\right)$ correspond to the values of the witness with rank-1 measurement $\Pi_{\boldsymbol{i}}^{\mathcal{SF}}$ in the computational basis $\left\{ \left|\boldsymbol{i}\right\rangle \right\} _{\boldsymbol{i}}$. The probability $p$ is the additional noise (either due to mixing with the noisy state or local depolarisation) on the initial state. \label{fig:Numerical-Simulation-ISBS}}
\end{figure}

Our exact numerical results are shown in Fig.~\ref{fig:Numerical-Simulation-ISBS}. The final unitary before measurement consists of Hadamards on all system-environment photons. Such a unitary gives a better lower bound to the witness than the Hadamard arrangement used in Fig.  \ref{fig:Experiment-circuit} for strong quantum Darwinism. We see that for the mixed, reduced system-fragment states, the witness is tight with the basis-dependent measure. It is not tight for the full system-environment state, which contains global quantum correlations, but it nonetheless  successfully witnesses non-objectivity in the state.

This witness scheme scales extremely well due to the lack of CNOT operations, requiring only $C+C$ total runs versus the $C\cdot3^{1+F}$ runs for $F$ photon environments in the case of quantum state tomography (cf. with subsec. \ref{subsec:Comparison-with-State-Tomography}). Here, one could potentially afford an entangling unitary in order to witness the quantum correlations in the full system-environment state.

\phantomsection
\addcontentsline{toc}{section}{References}

\bibliographystyle{apsrev4-1}
\bibliography{biblio}

\end{document}